\documentclass{article}
\usepackage[utf8]{inputenc}
\usepackage{url}
\usepackage{hyperref}
\usepackage{graphicx,amssymb,mathrsfs,amsmath,color,fancyhdr}
\usepackage{amsthm}
\usepackage{manfnt}
\usepackage{cases}
\usepackage{mathbbol}
\usepackage{geometry}
\usepackage{physics}
\usepackage{framed}
\usepackage{bm}
\usepackage{verbatim}

\usepackage[T1]{fontenc}
\usepackage{authblk}
\usepackage{geometry}
\usepackage{amsfonts}
\usepackage{graphicx,amssymb,mathrsfs,amsmath,color,fancyhdr}
\usepackage{subcaption}
\usepackage{amsthm}
\usepackage{manfnt}
\usepackage{cases}
\usepackage{mathbbol}
\usepackage{bm}
\usepackage{setspace}
\usepackage{hyperref}

\newcommand{\ddd}{\text{d}}

\newcommand{\keywords}[1]{\par\noindent\textbf{Keywords:} #1}

\newtheorem{proposition}{Proposition}
\newtheorem{theorem}{Theorem}

\newtheorem{example}{Example}
\newtheorem{lemma}{Lemma}
\newtheorem{remark}{Remark}

\allowdisplaybreaks

\title{Leveraging IS and TC: Optimal order execution subject to reference strategies}

\author[1]{Xue Cheng}
\author[1]{Peng Guo}
\affil[1]{\footnotesize LMEQF, Department of Financial Mathematics, School of Mathematical Sciences, Peking University, Beijing 100871, China.}

\author[2]{Tai-Ho Wang}
\affil[2]{\footnotesize Department of Mathematics,  Baruch College, CUNY, 1 Bernard Baruch Way, New York, NY 10010, USA }

\date{\today}

\begin{document}

\maketitle

\begin{abstract}
The paper addresses the problem of meta order execution from a broker-dealer's point of view in Almgren-Chriss model under execution risk. A broker-dealer agency is authorized to execute an order of trading on some client's behalf. The strategies that the agent is allowed to deploy is subject to a benchmark, referred to as the reference strategy, regulated by the client. We formulate the broker's problem as a utility maximization problem in which the broker seeks to maximize his utility of excess profit-and-loss at the execution horizon, of which optimal feedback strategies are obtained in closed form. In the absence of execution risk, the optimal strategies subject to reference strategies are deterministic. We establish an affine structure among the trading trajectories under optimal strategies subject to general reference strategies using implementation shortfall (IS) and target close (TC) orders as basis. Furthermore, an approximation theorem is proposed to show that with small error, general reference strategies can be approximated by piece-wise constant ones, of which the optimal strategy is piece-wise linear combination between IS and TC orders. We conclude the paper with numerical experiments illustrating the trading trajectories as well as histograms of terminal wealth and utility at investment horizon under optimal strategies versus those under TWAP strategies.
    
\end{abstract}


\tableofcontents

\newpage

\section{Introduction} \label{Sec:Intro}

    In the financial industry, large position holders such as pension funds or investment banks for various reasons are required to trade in or trade out from their current position to an updated target, possibly subject to a given execution horizon which may vary from days to weeks. The net holdings to be adjusted between the current and the target positions are usually too large to be simply dumped to the market without a priori deliberately assessing the trade's market impact. Untamed price impact by trading may result in significant transaction cost, potentially turned into a substantial loss. Such an execution risk requires to be properly managed and controlled; otherwise it would eventually influence the position holder's overall profit and loss (P\&L). A common practice is to delegate the execution to the firm's order execution department or outsource to a broker-dealer agency.  

%
    
The large position holders, while delegating their tasks to an agency for execution, may have in mind their own preferred strategies or benchmarks that they would like the delegated agent to closely track along.    
For instance, {\it implementation shortfall} (IS) orders \cite{almgren2001optimal} are frequently employed by managers for the purpose of short-term alpha pursuit. IS orders are constructed with a pre-trade benchmark price in mind, aiming at executing orders at an average price that remains relatively close to the market price at the beginning of the trade. Managers can often use arrival prices to help measure total trading costs: the closer the execution price is to the arrival price, the lower the associated costs (see CFA-level-II \cite{cfa2017cfa}). On the contrary, {\it target close} (TC) orders \cite{gueanttarget}, often deployed by index-fund managers for the purpose of minimizing fund risk and tracking error, are formulated with a post-trade benchmark price in order to secure a price in average that remains relatively close to the closing price. This is often important for mutual fund managers who manage funds that only calculate NAV once daily at closing. {\it Volume weighted average price} (VWAP) and {\it Time weighted average price} (TWAP) strategies are benchmarks specified for trading sequences with constant trading rate in wall clock time (TWAP) and in volume time (VWAP) respectively. These algorithmic trading strategies are examples of benchmarks that may be imposed as trading constraints to the agent who is missioned to trade in or trade out the position. 
We shall delve into this type of order execution problems subject to a pre-specified benchmark strategy, which we refer to as the {\it reference strategy} (RS for short), as a stochastic control problem and determine their corresponding optimal strategies in feedback form.

    The pioneering works in \cite{almgren1999value}, \cite{almgren2001optimal}, \cite{doi:10.1080/135048602100056}  and \cite{bertsimas1998optimal} are among the first to deal with the problem of order execution under price impact.
Since its introduction to the order execution problem, numerous progresses and extensions on the classical Almgren-Chriss framework have been made extensively. For instance, \cite{gatheral2010no} and \cite{gatheral2012transient} introduced the notion of transient impact to account for the dissipation of price impacts from the past trades. 
\cite{tse2013comparison} discusses the objective in the optimization problem, while \cite{cheng2019optimal} extends the classical mean-variance framework first considered in \cite{almgren1999value} to encompass general risk measures as penalty for risk aversion. \cite{frei2015optimal}, \cite{cartea2016closed}, \cite{gueant2014vwap} solve the optimal execution problem in relation to a VWAP benchmark, while \cite{almgren2007adaptive} asserts the use of the arrival price as a benchmark within the Almgren-Chriss framework. In fact, the arrival price, also known as ``pre-trade benchmark'', appears to be the most commonly used benchmark in academic papers. The closing price is the value of a security at its last transaction during a trading session \cite{gueanttarget} \cite{frei2018optimal}. Average price benchmarks may also appear in accelerated share repurchase (ASR) contracts \cite{gueant2015accelerated}\cite{bargeron2011accelerated}, which can be regarded as optimal execution problems with optimal stopping. Above all, \cite{cheng2017optimal} introduces the concept of execution risk, that pre-scheduled orders may not be fully executed, of which the empirical evidence is confirmed by \cite{carmona2023optimal} that the inventory processes of traders invariably contain a Brownian motion term, which contradicts the common assumption adopted in most optimal execution models that the inventory process is absolute continuous. Another perspective on introducing the noise term into the inventory process is that when a centralized trading desk aggregates order flows within a financial institution, stochastic order flow arises \cite{nutz2023unwinding} \cite{cartea2022brokers}. The aforementioned papers are by no means meant for an exhausting list in literature on this line of active research. 
    
    
    In the current paper, we consider the order execution problem from a broker-dealer's point of view. Assume that a broker is delegated to reallocate a client's holdings of a certain stock under the Almgren-Chriss model with execution risk. The broker is regulated by his client to track a benchmark strategy, i.e., the reference strategy, to the client's preference. The broker's incentive of executing client's order in this circumstance is to maximize his own expected P\&L excess to that of the reference strategy, marked-to-market. To account for risk aversion, we recast the broker's order execution problem as a utility maximization problem and, in certain cases, are able to solve the problem in closed form. In particular, when execution risk vanishes, or becomes negligible, we show that there exists an ``affine structure'' among the optimal strategies induced from various reference strategies. This algebraic structure is supposed to help the broker for concocting and understanding optimal strategies subject to client's general reference strategies. We argue that the framework is highly versatile in the sense that it encompasses commonly deployed execution strategies such as IS, TC as well as TWAP and VWAP orders as special cases for benchmarking. 
As per the algebraic structure, it follows that, for any given continuous reference strategy which can be approximated by a piece-wise constant function, we show that its corresponding optimal strategy can also be approximated by those induced from the piece-wise constant strategies.

    The rest of the paper is organized as follows. In Section \ref{sec:2}, we lay out the price impact model of Almgren-Chriss under execution risk and incorporate reference strategies into the problem of order execution. Section \ref{sec:3} presents the optimal feedback control of the order execution problem in Theorem \ref{thm:sto_ctrl} as one of the main results in the paper. Section \ref{sec:4} focuses and provides detailed discussions on the optimal strategies when execution risk vanishes. Reference strategies and their associated optimal strategies considered in Section \ref{sec:4} include IS and TC orders as well as piece-wise constant strategies. The emphasis is put on an ``affine structure'' among the trading trajectories induced by general reference strategies using unit IS and unit TC orders as a basis. 
Numerical examples illustrating the trading trajectories and the performance analysis under the optimal and TWAP strategies are shown and discussed in Section \ref{sec:numerics}. 
For the sake of smooth reading, technical proofs of all the theorems, propositions and lemmas are postponed and collected till the end of the paper as an appendix in Section \ref{sec:Appendix}.
    
    Throughout the paper, $\left(\Omega, \mathcal{F}, \mathbb{P}\right)$ denotes a complete probability space equipped with a filtration describing the information structure $\mathbb{F} := \left\{\mathcal{F}_{t}\right\}_{t \in [0,T]}$, where $t$ is the time variable and $T>0$ the fixed finite liquidation horizon. Let $\left\{W_t, Z_t\right\}_{t \in [0,T]}$ be a two-dimensional Brownian motion with constant correlation $\rho$ defined on $\left(\Omega, \mathcal{F}, \mathbb{P}\right)$. The filtration $\mathbb{F}$  is generated by the trajectories of the above Brownian motion, completed with all $\mathbb{P}$-null measure sets of $\mathcal{F}$.

\section{Model setup} \label{sec:2}

\subsection{Price impact model}

Assume a broker is delegated to reallocate a client's holdings of a certain stock from $x_0$ shares to $A$ shares within a given horizon $T$. Let $x_t$ be the number of shares the broker holds at time $t \in [0, T]$ during the reallocation process and $\widetilde{S}_t$ the transacted price at time $t$. The price dynamic is assumed to follow the Almgren–Chriss model \cite{almgren1999value} \cite{almgren2001optimal} \cite{doi:10.1080/135048602100056}. In the Almgren–Chriss framework, the transacted price $\widetilde{S}_t$ consists of the fair price $S_t$ and a slippage. The fair price $S_t$ is driven by the SDE
\[
    \ddd S_t = \mu \ddd t+ \gamma \ddd x_t +\sigma \ddd W_t,
\]
or equivalently,
\[
    S_t = S_0 + \mu t+ \gamma(x_t-x_0)+\sigma W_t,
\]
where $\mu t$ describes the tendency of the stock and $(W_t)_{t\in [0,T]}$ is a standard Brownian motion. The term $\gamma(x_t-x_0)$, for $\gamma\ge 0$, is usually referred to as the \textit{permanent impact}. Penalized by a price slippage, the transacted price is thus given by
\[
    \widetilde{S}_t = S_t - \eta v_t,
\]
where $v_t$ denotes the broker's intended trading rate at the instant $t$. The slippage $-\eta v_t$, for $\eta \ge 0$, is also referred to as the \textit{temporary impact}. We remark that in the original setting of Almgren-Chriss and its extensions, the trading rate $v_t$ plays a dual role. On the one hand, it serves as the realized trading rate per the relationship $v_t=-\dot{x_t}$ between $x_t$ and $v_t$. On the other hand, it is regarded as the control variable in the problem of optimal execution. These two seemingly distinct roles coincide if all the scheduled orders are guaranteed fully executed. However, in reality it is well noticed among practitioners that, while executing a sequence of pre-scheduled orders, the orders in the sequence may not be fully executed, resulting in an uncontrollable realized order flow. This introduces an additional risk to the order execution problem, the \textit{execution risk}. To account for this uncertainty, we introduce a noise component driven by a correlated Brownian motion $Z_t$ into the dynamic of the position $x_t$ as
\[
    \ddd x_t = -v_t \ddd t+m(v_t)\ddd Z_t.
\]
The diffusion term $m(v)$ characterizes the magnitude of execution risk. It is worth to reiterate that in a recent work in \cite{carmona2023optimal}, the authors showed that the presence of a Brownian component in the broker's inventory during reallocation process is statistically significant. 
Moreover, because of this execution risk, the broker is no longer guaranteed to achieve his intended position at the terminal time $T$, which gives rise to an additional opportunity cost. As a result, the broker is obligated to take a final block trade at time $T$ at
a worse price. Overall, the realised P\&L at the horizon $T$ is given by
\[
    \Pi:= (x_T-A)S_T-\beta(x_T-A)^2+\int_0^T\left(-\tilde{S_t}\right)\ddd x_t,
\]
where the term $\beta(x_T-A)^2$, for $\beta \ge 0$, penalizes the discrepency from a final block trade.

The following proposition shows that the P\&L can be written as an It\^{o} process. 
\begin{proposition} \label{prop:PnL}
    The realised P\&L can be rewritten as
\[
\begin{aligned}
    \Pi =& (x_0-A)S_0-\beta(x_T-A)^2 +\int_0^T \left(\mu (x_t-A)+\rho \sigma m(v_t)-\eta v_t^2-\gamma v_t(x_t-A) \right) \ddd t \\
    &+ \int_0^T \sigma (x_t-A) \ddd W_t + \int_0^T (\eta v_t+\gamma (x_t-A))m(v_t)\ddd Z_t.
\end{aligned}
\]
\end{proposition}


\subsection{Reference strategy} \label{sec:2.2}

A common practice in order execution brokerage is that clients may come forward to brokers with their own preferred strategies for benchmarking.   
These benchmark strategies can be either strategies suggested by elite investors or commonly used ones such as TWAP strategies. We shall refer to these pre-specified benchmark strategies as \textit{reference strategies}. 
We consider the reference strategies that can be represented by a deterministic function $(R_t)_{t\in [0,T]}$ with $R_0\equiv x_0$ and $R_T\equiv A$ hereafter. Moreover, we assume that $R_t$ is a differentiable a.e. function. The broker's incentive of executing client's order is thus to maximize his own expected P\&L excess to that of the reference strategy, marked-to-market. Specifically, by disregarding the price impact and slippages incurred from order execution, the stock price $S_t^0$ reads
\[
    S_t^{0} = S_0 + \mu t + \sigma W_t.
\]
The marked-to-market P\&L $\Pi^R$ for the reference strategy $R$ is evaluated as      
$$
\begin{aligned}
   \Pi^R :=& \int_0^T \left(-S_t^0\right) \ddd R_t \\
   =& -(S_t^0R_t)\big|_0^T + \int_0^T R_t (\mu \ddd t+\sigma \ddd W_t) \\
   =& -AS_T^0 + x_0S_0 + \int_0^T \mu R_t \ddd t + \int_0^T \sigma R_t \ddd W_t \\
   =& (x_0-A)S_0 + \int_0^T \mu (R_t-A) \ddd t + \int_0^T \sigma (R_t-A) \ddd W_t.
\end{aligned}
$$
Hence, the broker's excess P\&L $\tilde\Pi$, defined by the difference between $\Pi$ and $\Pi^R$, is given by 
\begin{equation}
\begin{aligned}
    \tilde{\Pi} :=& \Pi-\Pi^R \\
    =& -\beta (x_T-A)^2+\int_0^T \left(\mu (x_t-R_t)+\rho \sigma m(v_t)-\eta v_t^2-\gamma v_t(x_t-A) \right) \ddd t\\
    &+ \int_0^T \sigma (x_t-R_t) \ddd W_t + \int_0^T (\eta v_t+\gamma (x_t-A))m(v_t)\ddd Z_t.
\end{aligned}
\end{equation}
The broker's goal is thus to maximize his expected excess P\&L in a risk aversion manner which we recast as a utility maximization problem in the section that follows.

\section{Optimal execution as utility maximizing} \label{sec:3}

In this section, we recast the problem of order execution as a utility maximization problem as follows. Recall that $(\Omega, \mathcal{F}, P)$ denotes a complete probability space equipped with a filtration
${(\mathcal{F}_t)}_{t\ge 0}$ satisfying the usual conditions. We assume that all random variables and stochastic processes are defined on $(\Omega, \mathcal{F},
{(\mathcal{F}_t)}_{t\ge 0}, \mathbb{P})$. The set of all real-valued progressively measurable processes are denote by $\mathcal{M}$, while the collection of admissible controls $\mathcal{A}$ is set as
\[
    \mathcal{A} := \left\{ v\in \mathcal{M}: \int_0^T \mathbb{E}[v_t^2] \ddd t < \infty \right\}.
\]
to meet the assumptions given in \cite{yong2012stochastic} and \cite{duncan2013linear}, which assure the solvability of both the risk-neutral problem and the risk-aversion one. The broker's problem is to determine an optimal admissible strategy $v^*$ that maximizes his expected exponential utility at the horizon $T$, i.e., 
\begin{equation}
\begin{aligned}
   & \sup_{v\in \mathcal{A}} \mathbb{E} \left[ u\left( \tilde{\Pi} \right) \right],
\end{aligned}
    \label{sto_ctrl}
\end{equation}
where the utility function $u(x):= \frac{1}{\theta}(1-e^{-\theta x}), \theta > 0$ represents the broker's preference following CARA (Constant Absolute Risk Aversion) preference and $\theta > 0$ is the risk-aversion parameter. Note that as $\theta \downarrow 0$, the utility function $u(x)=\lim_{\theta \downarrow 0}\frac{1}{\theta}(1-e^{-\theta x}) = x$ reflects a risk-neutral preference. When the utility function is selected as $u(x)=x$, we solve the stochastic control problem and give the close form of the value function $\mathbb{E}\left[  \tilde{\Pi} \right]\bigg|_{v=v^{**}} = \sup_{v\in \mathcal{A}} \mathbb{E}\left[  \tilde{\Pi} \right]< \infty$, where $v^{**} \in \mathcal{A}$ is the optimal feedback control in the risk-neutral case, see in Appendix \ref{Appendix:risk-neutral}. Moreover, when $u(x)=\frac{1}{\theta}(1-e^{-\theta x}), u'(x)=e^{-\theta x}>0, u''(x)=-\theta e^{-\theta x} < 0$, so $u(\cdot)$ is a increasing concave function. For any admissible control $\tilde{v}$, by Jensen's inequality,
\[
    \mathbb{E} \left[ u\left( \tilde{\Pi} \right) \right]\bigg|_{v=\tilde{v}} \le u\left(\mathbb{E} \left[  \tilde{\Pi} \right]\bigg|_{v=\tilde{v}} \right) \le u\left(\mathbb{E} \left[  \tilde{\Pi} \right]\bigg|_{v=v^{**}} \right) < \infty,
\]
which shows the finiteness of the utility maximizing problem. In the following, we solve the utility maximization problem \eqref{sto_ctrl} and present solutions in closed form in the cases where the execution risk $m(v)$ is either a constant or zero. The case of zero execution risk is postponed and will be discussed in more details in Section \ref{sec:4}.

\subsection{Optimal execution with risk aversion}

When the execution risk is constant, i.e., $m(v)\equiv m_0$ for some fixed constant $m_0$, the utility maximizing problem reduces to the following Stochastic Linear Exponential Quadratic (SLEQ) control problem \cite{duncan2013linear} \cite{lim2005new}
\begin{equation}
    \begin{cases}
    \begin{aligned}
        \sup_{v \in \mathcal{A}}  \mathbb{E} \left[ u\left( \tilde{\Pi} \right) \right] = &  \sup_{v \in \mathcal{A}} \begin{aligned}[t]
                 \mathbb{E} \Bigg[ \frac{1}{\theta} \Bigg( 1-\exp\Bigg( &\theta \beta (x_T-A)^2-\int_0^T \theta \left(\mu (x_t-R_t)+\rho \sigma m_0 -\eta v_t^2-\gamma v_t(x_t-A) \right) \dd t \\
              -& \int_0^T \theta\sigma (x_t-R_t) \dd W_t - \int_0^T \theta(\eta v_t+\gamma (x_t-A))m_0\dd Z_t \Bigg)\Bigg)   \Bigg], 
              \end{aligned}\\
        & \text{s.t. }   x_t = x_0 - \int_0^t v_s \dd s + m_0 Z_t.
    \end{aligned}
    \end{cases}
\end{equation}
Define the value function of the problem by
\begin{equation} \label{eq:value}
         \begin{aligned} 
              V(t,x):= \sup_{v \in \mathcal{A}}  \mathbb{E} \Bigg[ \frac{1}{\theta} \Bigg( 1-\exp\Bigg( &\theta \beta (x_T-A)^2-\int_t^T \theta \left(\mu (x_s-R_s)+\rho \sigma m_0 -\eta v_s^2-\gamma v_s(x_s-A) \right) \dd s \\
              -& \int_t^T \theta\sigma (x_s-R_s) \dd W_s - \int_t^T \theta(\eta v_s+\gamma (x_s-A))m_0\dd Z_s \Bigg)\Bigg) \Bigg | x_t=x  \Bigg], 
              \end{aligned}
\end{equation}
where $\mathcal{A}_t:=\left\{ v \text{ is progressively measurable in } [t,T] \text{ and } \int_t^T \mathbb{E}[v_s^2] \dd s < \infty \right\}$. The optimal feedback control can be obtained in closed form. We summarize the result in the following theorem.

\begin{theorem} \label{thm:sto_ctrl}
    The value function (\ref{eq:value}) of the utility maximization problem has the closed form expression
    \[
        V(t,x) = \frac{1}{\theta} \left( 1 - \exp \left\{ \left( b_2(t)+\frac{\theta\gamma}{2} \right) (x-A)^2 + b_1(t)(x-A) + b_0(t) \right\} \right).
    \]
    where the parameters $H$, $l_1$, and $l_3$ are given by
$$
\begin{cases}
\begin{aligned}
H=& \ \theta \eta + \frac{1}{2}\theta^2 \eta^2 m_0^2,\\
l_3=& \ \frac{1}{2}m_0\eta \theta^2 \rho \sigma, \\
l_1=& \ \frac{\theta^2 \sigma^2}{2}H.
\end{aligned}
\end{cases}
$$
The time dependent function $b_2$ is given in closed form by 
$$
    b_2(t) := \sqrt{l_1}\coth \left( A_0 + \frac{\sqrt{l_1}}{H}(T-t) \right)-l_3, \ \ \text{where} \ \ A_0 := \coth^{-1} \left( \frac{l_3+\frac{\theta}{2}(2\beta-\gamma)}{\sqrt{l_1}} \right),
$$
$b_1$ is the solution to the following terminal value problem 
$$
    b_1' (t)= \theta \mu + \frac{1}{H}b_1(t)\Big(b_2(t)+l_3\Big) + \frac{\eta \theta^2\sigma}{H}(R_t-A)\left( \theta \sigma + \frac{1}{2}m_0^2\theta^2\eta \sigma (1-\rho^2)-\rho m_0 b_2(t) \right), \ \ b_1(T) = 0,
$$
and
\[
    b_0(t) := \int_t^T \left[ \frac{l_1-l_3^2}{H} (R_s-A)^2 + \left( \theta \mu + \frac{l_3}{H}b_1 \right)(R_s-A) + \left( -\frac{b_1^2}{4H}+m_0^2\left( b_2+\frac{\theta \gamma}{2} \right)-\rho m_0\theta \sigma \right) \right] \dd s.
\]
Moreover, the optimal feedback control $v^* \in \mathcal{A}$ of the utility maximization problem (\ref{sto_ctrl}) is given by
\begin{equation}
    v^*_t = \frac{(1+\theta \eta m_0^2)b_2(t)}{H} \cdot (x_t-A) + \frac{1+\theta \eta m_0^2}{2H} \cdot b_1(t) + \frac{l_3}{H}\cdot (R_t-x_t), \label{eqn:opt-v}
\end{equation}

\end{theorem}

It's worth mentioning that this theorem does not contain much mathematical breakthrough, instead, we regard it as a fundamental result to introduce valuable results in section \ref{sec:4}, where the affine structure is what we find the most interesting.

\begin{remark} \label{remark:feedback_control}

The optimal feedback control $v_t^*$ in (\ref{eqn:opt-v}) consists of three parts:
\begin{itemize}
    \item The first part $\frac{(1+\theta \eta m_0^2)b_2}{H} \cdot (x_t-A)$ has the same sign as $x_t-A$. Without loss of generality, we assume that $x_0>A$, when $x_t\ge A$, which implies the reallocation process has not finished, so the broker should continue to sell the stock. Conversely, when $x_t < A$, which implies the strategy is over-shooting, so the broker should buy some shares of the stock back.

    \item The second part $\frac{1+\theta \eta m_0^2}{2H} \cdot b_1$, which is a pre-specified deterministic function in time $t$, does not depend on the state variable $x_t$.

    \item The third part $\frac{l_3}{H}\cdot (R_t-x_t)$ has the same sign as $\rho(R_t-x_t)$, where $\rho$ is the correlation between the stock price process and the execution risk. The discussions in \cite{carmona2013self} state the fact that the sign of $\rho$ depends on the order type adopted by the trader due to adverse selection: when the trader uses market orders, the correlation $\rho$ is negative while $\rho$ is positive whenever trading with limit orders. In this paper, we regard $\rho$ as a market parameter reflecting whether the market is trader-friendly or not.
    
    \begin{itemize}
        \item When $\rho \ge 0$, the stock price and execution risk tend to increase or decrease together: when the stock price increases, execution risk makes traders buy more or sell less than the amount they submit, which is benefit to the traders; when the stock price decreases, execution risk forces traders to buy less or sell more, which is also benefit to the traders. In such trader-friendly market, like the figure (\ref{fig:rho_positive}), the broker should keep away from the reference strategy $R_t$ because an overly conservative strategy is not necessary. 


        \item On the contrary, $\rho < 0$ implies a trader-harmful market: when the stock price increases, execution risk makes traders buy less or sell more than the amount they submit, which is harm to the traders; when the stock price decreases, execution risk forces traders to buy more or sell less, which is also harm to the traders. In such trader-harmful market, like the figure (\ref{fig:rho_negative}), the broker should keep close to the reference strategy $R_t$ to attain lower risk.
\begin{figure*}[htbp]
    \centering
    \begin{subfigure}[t]{0.5\textwidth}
        \centering
        \includegraphics[height=1.2in]{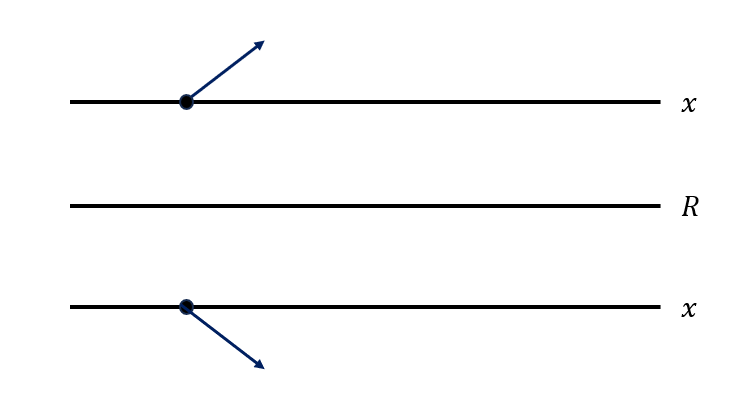}
        \caption{A trader-friendly market}
        \label{fig:rho_positive}
    \end{subfigure}%
    ~ 
    \begin{subfigure}[t]{0.5\textwidth}
        \centering
        \includegraphics[height=1.2in]{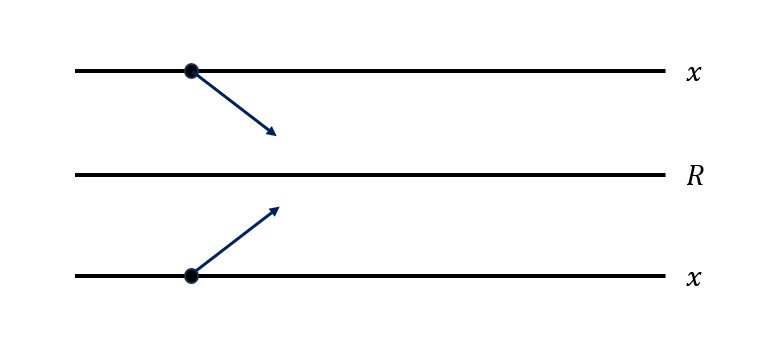}
        \caption{A trader-harmful market}
        \label{fig:rho_negative}
    \end{subfigure}
    \caption{Market parameter $\rho$}
\end{figure*}
\end{itemize}

\end{itemize}

\end{remark}


\begin{remark}

By applying the optimal strategy (\ref{eqn:opt-v}), one obtains the optimal liquidation trajectory $x_t^*$ as follows. For simplicity, take the parameters $A=\rho=0$, $R_t \equiv 0$ and $\mu=0$, $x_t^*$ satisfies the SDE
\[
    \dd x_t^* = - \left( \frac{(1+\theta \eta m_0^2)b_2}{H}x^*_t + \frac{(1+\theta \eta m_0^2)b_1}{2H} \right) \dd t + m_0 \dd Z_t,
\]
as $\beta \to +\infty$, $x_t^*$ has the limit
\[
    x_t^{**}= \left( \frac{\sinh \left(\frac{\sqrt{l_1}}{H}(T-t)\right)}{\sinh \left(\frac{\sqrt{l_1}}{H}T\right)} \right)^{\frac{2H}{\theta \eta}-1} x_0 + m_0 \int_0^t \left( \frac{\sinh \left(\frac{\sqrt{l_1}}{H}(T-t)\right)}{\sinh \left(\frac{\sqrt{l_1}}{H}(T-s)\right)} \right)^{\frac{2H}{\theta \eta}-1} \dd Z_s.
\]
Furthermore, when $m_0\to 0$, $H\to \theta \eta$, the limit of the expression above is
\[
    \lim_{m_0\to 0} x_t^{**} = \frac{\sinh(\kappa(T-t))}{\sinh (\kappa T)} x_0,
\]
where $\kappa=\sqrt{\frac{\theta\sigma^2}{2\eta}}$. It recovers the classical implementation shortfall (IS) strategy (see in section \ref{subsec:4.2}), which implies the fact that our strategy can be regarded as a version of adaptive IS strategy. Furthermore, we may find that
\[
    \lim_{m_0\to 0} \frac{x_t^{**} - \frac{\sinh(\kappa(T-t))}{\sinh (\kappa T)} x_0}{m_0} = \int_0^t \frac{\sinh \left(\kappa(T-t)\right)}{\sinh \left(\kappa(T-s)\right)}  \dd Z_s,
\]
which is an Ornstein-Uhlenbeck bridge \cite{mazzolo2017constraint}. Therefore, when $m_0$ is small, one may find an approximation to the optimal strategy
\[
    x_t^{**} \approx \frac{\sinh(\kappa(T-t))}{\sinh (\kappa T)} x_0 + m_0 \int_0^t \frac{\sinh \left(\kappa(T-t)\right)}{\sinh \left(\kappa(T-s)\right)}  \dd Z_s,
\]
which is a classical IS strategy plus an Ornstein-Uhlenbeck bridge, which reduces the strategy's inventory risk. Figure (\ref{fig:inventory_risk}) shows the inventory risk of the optimal strategy and the classical IS strategy, which shows the fact that our optimal strategy has lower inventory risk than the classical IS strategy, especially at the end of the trading process.

\begin{figure}[htbp]
    \centering
	\includegraphics[scale=0.4]{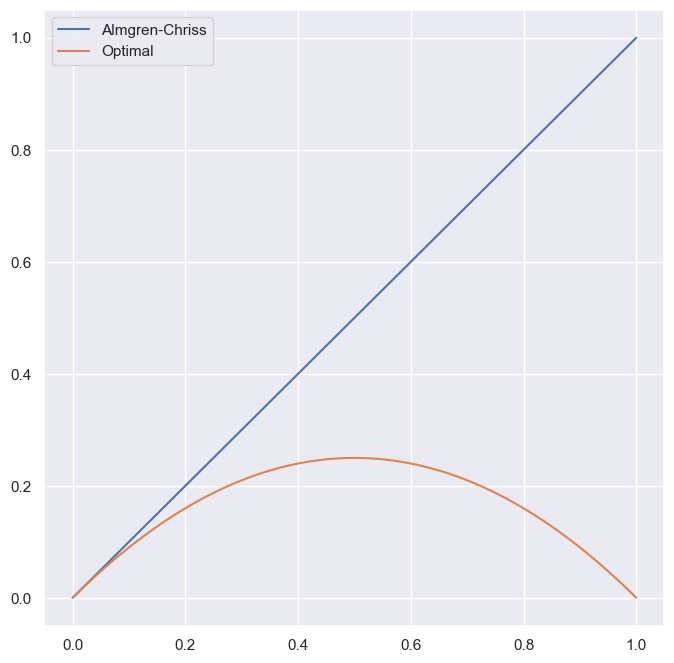}
	\caption{Inventory risk of the optimal strategy and the classical IS strategy}
	\label{fig:inventory_risk}
\end{figure}

\end{remark} 

To summary the section, it should be emphasized that despite the optimal feedback control behaves linearly in the states carries a certain meaning which we discuss in Remark \ref{remark:feedback_control}, however, the case which we care most is where the execution risk does not exist, which is a limit case of Theorem \ref{thm:sto_ctrl}, discussed in the next section.

\section{Zero execution risk} \label{sec:4}
For the purpose of better understanding as well as visualizing the optimal strategy obtained in \eqref{eqn:opt-v}, we present in this section the trading trajectory $x_t$ under optimal strategy in the case when $m_0 = 0$ for various reference strategies. In Section \ref{subsec:4.1}, we give the optimal strategy for general reference strategies. In Section \ref{subsec:4.2}, we present two classic trading strategies: IS order and TC order, both widely used in the market, and show that they are special cases of our model. Next, we consider the case when the reference strategy is an endpoints-only one in Section \ref{subsec:4.4}, of which the optimal strategy has a heuristic form. Last but not least, piece-wise constant reference strategies are studied in Section \ref{sec:5}.

Notice that in this case the setting of the problem reduces to that of the Almgren-Chriss but subject to a reference strategy. Furthermore, we shall set the final penalty parameter $\beta \to +\infty$, indicating that a final block trade at terminal time $T$ is strictly prohibited.

\subsection{General reference strategy} \label{subsec:4.1}
For a given generic reference strategy $R_t$, the following theorem shows a representation for the trading trajectory $x_t$ under optimal control $v^*$.  

\begin{theorem} \label{thm:det_ctrl}
    Let $m(v) \equiv 0$ and $\beta \to +\infty$. The trajectory $x_t$ under the solution to the optimization problem (\ref{sto_ctrl}) is given by
    \[
    \begin{aligned}
    x_t =& \frac{\kappa}{\sinh \kappa T} \int_0^T \bigg ( R_s  \sinh (\kappa \min(s,t)) \sinh (\kappa (T-\max(s,t))) \bigg) \ddd s \\
    &+ \left( x_0 -\frac{\mu}{\theta \sigma^2} \right) \cdot \frac{\sinh \kappa(T-t)}{\sinh \kappa T} + A + \left( -A + \frac{\mu}{\theta \sigma^2}\right) \cdot \frac{\sinh \kappa T - \sinh \kappa t}{\sinh \kappa T},
    \label{solution}
\end{aligned}
\]
with the tuning parameter $\kappa := \sqrt{\frac{\theta \sigma^2}{2\eta}}$.
\end{theorem}
A few remarks on the trajectory $x_t$ shown in Theorem \ref{thm:det_ctrl} are in order. 
\begin{remark}
    The parameter $\kappa$ can be regarded as a tuning parameter between the TWAP and the reference strategies if $\mu = 0$. Specifically, in the limit as $\kappa$ approaches zero, the optimal strategy converges to the TWAP strategy, i.e., for any $t \in [0, T]$, we have 
\[
    \lim_{\kappa \to 0} x_t = A + (x_0-A) \cdot \left( 1-\frac{t}{T} \right).
\]
On the other extreme as $\kappa \to \infty$, the optimal trajectory converges to the reference strategy $R_t$ itself
\[
    \lim_{\kappa\to\infty}x_t = R_t
\]
for any $t \in [0, T]$.
\end{remark}

\begin{remark}
    The optimal trajectory may also be written as an ``affine transformation'' as 
\[
  x_t= \frac{\mu}{\theta \sigma^2} + a_t \cdot \frac{\sinh\kappa(T-t)}{\sinh \kappa T} + b_t \cdot \frac{\sinh \kappa t}{\sinh \kappa T},
\]
where
\begin{eqnarray*}
&& a_t = x_0 - \frac{\mu}{\theta \sigma^2} + \int_0^t \kappa R_s \sinh \kappa s \ddd s, \\ 
&& b_t = A - \frac{\mu}{\theta \sigma^2} + \int_t^T \kappa R_{s} \sinh \kappa (T-s) \ddd s.
\end{eqnarray*}
A more detailed discussion on this affine transformation can be found in Section \ref{sec:5}.
\end{remark}

Next, we further specialize the reference strategies and present their corresponding trajectories in the following sections. 

\subsection{Implementation shortfall (IS) order and target close (TC) order} \label{subsec:4.2}

Assume $\mu=0$ and the broker is asked to take as the reference strategy a block trade of size $|A - x_0|$ at $t=0$, i.e.,
\[
\begin{aligned}
    R_t^{\text{IS}}=
    \begin{cases}
        x_0, \ & t=0,\\
        A, \ & 0<t\le T.
    \end{cases}
\end{aligned}
\]
The corresponding marked-to-market P\&L of the reference strategy $R^{\rm IS}$ is given by
\[
\begin{aligned}
    \Pi^{\text{IS}} := \int_0^T \left(- S_t^{\text{IS}} \right) \ddd R_t^{\text{IS}} =& -(x_0-A)\cdot \frac{S_{0-}^{\text{IS}}+S_{0+}^{\text{IS}}}{2} \\
    =& -(x_0-A)S_0+\frac{\gamma}{2}(x_0-A)^2,
\end{aligned}
\]
which is equivalent to choosing the initial price $S_0$ as reference price. This model is referred to as an \textit{implementation shortfall model} in \cite{gueant2016financial}, of which the optimal strategy is 
\begin{equation}
x_t=A+(x_0-A)\cdot \text{IS}_t, \label{eqn:opt-x-is}
\end{equation}
where
\begin{equation}
    \text{IS}_t := \frac{\sinh \kappa (T-t)}{\sinh \kappa T}.
    \label{eq:IS}
\end{equation}
We shall refer to the strategy $\text{IS}_t$ as the {\it unit implementation shortfall (IS) order}. Hence, the trajectory in \eqref{eqn:opt-x-is} can be interpreted as executing $x_0-A$ unit IS orders if a broker is missioned to make transition of his position from $x_0$ to $A$ shares. We remark that, as we will show in Section \ref{sec:5}, the unit IS orders are used as one of the building blocks for the construction of optimal trajectories for more general reference strategies.

\begin{figure*}[htbp]
    \centering
    \begin{subfigure}[t]{0.48\textwidth}
        \centering
        \includegraphics[height=2.2in]{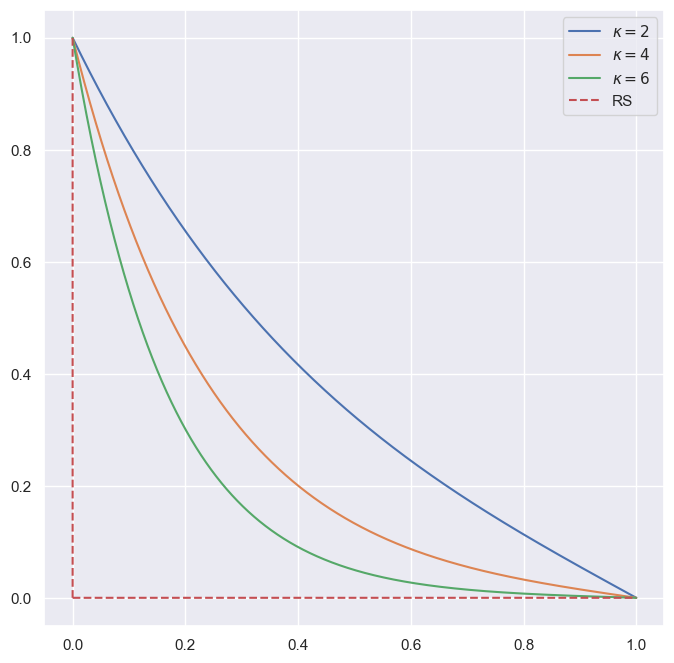}
        \caption{Unit implementation shortfall orders}
        \label{fig:IS}
    \end{subfigure}%
    ~ 
    \begin{subfigure}[t]{0.48\textwidth}
        \centering
        \includegraphics[height=2.2in]{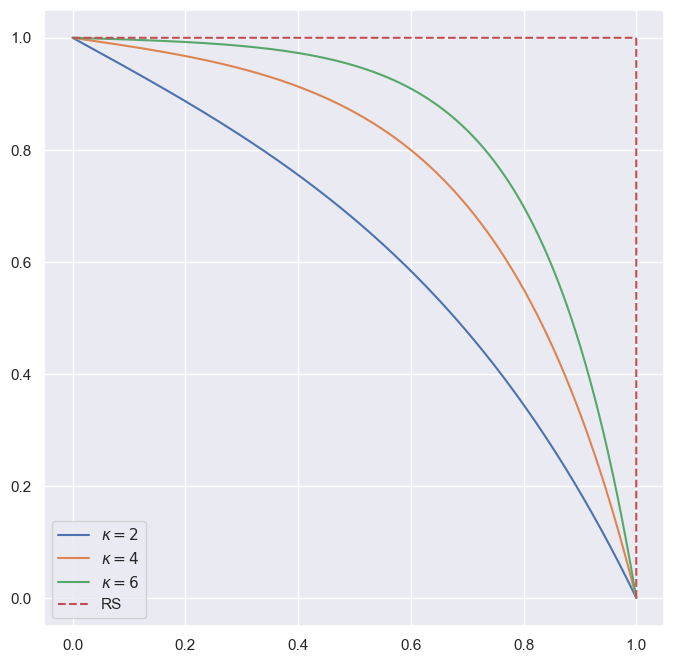}
        \caption{Unit target close orders}
        \label{fig:TC}
    \end{subfigure}
    \caption{Basic strategies in varying parameter $\kappa$.}
\end{figure*}

Figure (\ref{fig:IS}) shows examples of the unit implementation shortfall orders under various values of $\kappa$. Notice that, since $\kappa = \sqrt{\frac{\theta \sigma^2}{2\eta}}$, as $\theta, \sigma$ increase or $\eta$ decreases, thus $\kappa$ increases, the trajectories suggest the broker trade faster at the beginning and more slowly for the rest of the reallocation process. Indeed, the larger the $\kappa$, the faster the trading at the beginning. The financial rationale is as follows. Recall that $\theta$ is the coefficient of relative risk aversion for the utility function and $\sigma$ is the volatility of stock price. Hence, if either of the parameters increases, the broker is more concerned with the price risk than the execution risk, he tends toward trading faster at the beginning and more slowly thereafter. Also, since $\eta$ is the coefficient of slippage,  as it decreases, the slippage decreases as well, resulting in lower cost from faster transactions. Therefore, again to mitigate the concern of price risk during reallocation process, the broker should also trade faster at the beginning as well. 

Finally, we remark that, as $\kappa \to 0$, the unit IS order \eqref{eq:IS} converges to the \textbf{TWAP} (Time-Weight-Average-Price) strategy: for any $t\in [0,T]$,
\[
    \lim_{\kappa \to 0} \text{IS}_t = \lim_{\kappa \to 0} \frac{\sinh \kappa(T-t)}{\sinh \kappa T} = \frac{T-t}{T}.
\]
On the flip side, as $\kappa \to +\infty$, the optimal trajectory \eqref{eq:IS} converges to a block trade at time $0$:
\[
    \lim_{\kappa \to \infty} \text{IS}_t = \lim_{\kappa \to \infty} \frac{\sinh \kappa(T-t)}{\sinh \kappa T} = \lim_{\kappa \to \infty} e^{-\kappa t} \cdot \frac{1-e^{-2\kappa(T-t)}}{1-e^{-2\kappa T}}=
    \begin{cases}
        1, \ t=0,\\
        0, \ 0<t\le T.
    \end{cases}
\]
Assume again that $\mu=0$. Here the broker is given as the reference strategy to take only a block trade of size $|A - x_0|$ at terminal time $T$. That is,
\[
    R_t^{\text{TC}}=
    \begin{cases}
        x_0, \ 0\le t<T,\\
        A, t=T.
    \end{cases}
\]
The marked-to-market P\&L $\Pi^{\text{TC}}$ for this reference strategy is given by
\[
\begin{aligned}
    \Pi^{\text{TC}} = \int_0^T \left(- S_t^{\text{TC}} \right) \ddd R_t^{\text{TC}} =& -(x_0-A) \cdot \frac{S_{T-}^{\text{TC}}-S_{T+}^{\text{TC}}}{2} \\
    =& -(x_0-A)S_{T+}^{\text{TC}} - \frac{\gamma}{2}(x_0-A)^2,
\end{aligned}
\]
which is equivalent to choosing the stock price $S_T$ at terminal time as the reference price. This model is referred to as an \textit{target close model} \cite{gueant2016financial}\cite{gueanttarget}, of which the optimal strategy is given by 
\begin{equation}
x_t=A+(x_0-A)\cdot \text{TC}_t, \label{eqn:opt-x-tc}
\end{equation} 
where
\begin{equation}
    \text{TC}_t := \frac{\sinh \kappa T-\sinh \kappa t}{\sinh \kappa T}.
    \label{eq:TC}
\end{equation}
We refer to the strategy $\text{TC}_t$ in \eqref{eq:TC} as the {\it unit target close (TC) order}. 
Thus, similar to the case of implementation shortfall order in Section \ref{subsec:4.2}, the trajectory in \eqref{eqn:opt-x-tc} can be interpreted as executing $x_0-A$ unit TC orders if a broker is missioned to make transition of his position from $x_0$ to $A$ shares. Likewise as for the unit IS orders, we show in Section \ref{sec:5} that the unit TC orders are used as the other building blocks for the construction of optimal trajectories for more general reference strategies.

Figure (\ref{fig:TC}) shows examples of plots for unit target close orders in different values of $\kappa$. We observe that the trading trajectories are concave in time as opposed to convex in time as that of the unit IS orders. Also, as $\kappa$ increases, the optimal trajectory moves towards the reference strategy and towards the TWAP trajectory as $\kappa$ decreases. 
In fact, similar to the case for implementation shortfall, in the limit as $\kappa \to 0$, the unit TC order \eqref{eq:TC} converges to the TWAP strategy: for any $t\in [0,T]$,
\[
    \lim_{\kappa \to 0} \text{TC}_t = \lim_{\kappa \to 0} \frac{\sinh \kappa T-\sinh \kappa t}{\sinh \kappa T} = 1-\frac{t}{T};
\]
whereas in the other extreme as $\kappa \to +\infty$, the optimal trajectory converges to a block trade at terminal time $T$:
\[
    \lim_{\kappa \to \infty} \text{TC}_t = \lim_{\kappa \to \infty} \frac{\sinh \kappa T-\sinh \kappa t}{\sinh \kappa T} = 1-\lim_{\kappa \to \infty} e^{-\kappa (t-T)} \cdot \frac{1-e^{-\kappa t}}{1-e^{-\kappa T}}=
    \begin{cases}
        0, \ 0\le t<T,\\
        1, \ t=T.
    \end{cases}
\]

\subsection{Endpoints-only reference strategy} \label{subsec:4.4}
In this section, we consider the following reference strategy $R_t$, which we called the {\it endpoints-only}.
\begin{equation}
\label{eq:const_RS}
    R_t=
    \begin{cases}
        x_0, \ t=0,\\
        R, \ 0<t<T,\\
        A, \ t=T.
    \end{cases}
\end{equation}
The strategy is simply to hold $R$ shares during the entire reallocation process, except at the initial and terminal times,. 

The following lemma will prove itself useful in the determination of the optimal strategy subject to the reference strategy \eqref{eq:const_RS}. 
\begin{lemma} \label{lemma:integral}
We have that 
    \[
    \begin{aligned}
        & \frac{\kappa}{\sinh \kappa T} \int_0^T \bigg (\sinh (\kappa \min(s,t)) \sinh (\kappa (T-\max(s,t))) \bigg) \dd  s \\
        =& \frac{\sinh \kappa T-\sinh \kappa t}{\sinh \kappa T} - \frac{\sinh \kappa (T-t)}{\sinh \kappa T}    \\
        =& \text{TC}_t - \text{IS}_t.
    \end{aligned}
    \]
In other words, the given integral can be written as the difference between a unit TC and a unit IS order. 
\end{lemma}
We summarize the result for optimal strategies subject to \eqref{eq:const_RS} in the proposition that follows. 
\begin{proposition} \label{prop:const_RS}
With the reference strategy given in (\ref{eq:const_RS}), the optimal trajectory is given by an affine combination of a unit IS order and a unit TC order as 
    \begin{equation} \label{eqn:opt-x-const}
    x_t = \left( x_0-\frac{\mu}{\theta \sigma^2}-R \right)\cdot \text{IS}_t + \left( -A+ \frac{\mu}{\theta \sigma^2}+R \right) \cdot \text{TC}_t + A.
\end{equation}
Note that when $\mu=0$ and $R = x_0$, the optimal strategy reduces to the following TC strategy as in \eqref{eqn:opt-x-tc} 
\[
    x_t = (x_0-A) \cdot \text{TC}_t + A;
\]
whereas when $\mu=0$ and $R = A$, the optimal strategy reduces to the following IS strategy as in \eqref{eqn:opt-x-is}
\[
    x_t = (x_0-A) \cdot \text{IS}_t + A.
\]

\end{proposition}

\begin{figure}[htbp]
    \centering
	\includegraphics[scale=0.5]{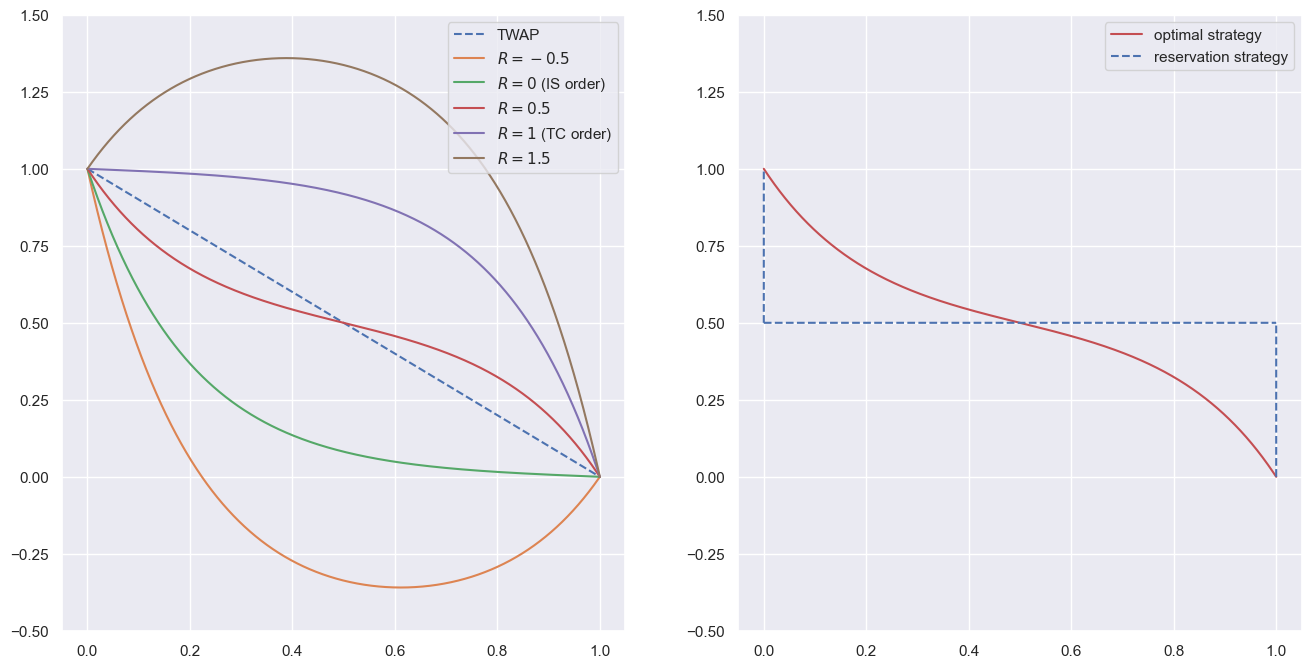}
	\caption{Optimal strategies in different reference levels $R$.}
	\label{fig:RP}
\end{figure}

Note that for any given reference level  $R$, the optimal trajectory in \eqref{eqn:opt-x-const} at any point in time is an affine combination of a unit IS order and a unit TC order. The coefficients represent a trade-off between the two strategies. For example, if $R$ is closer to the initial position $x_0$ and further away from the target position $A$, the strategy in \eqref{eqn:opt-x-const} suggests we weigh in more on the unit TC order and less on the unit IS order; conversely, one should then weigh in more on the unit IS order.   
%
%
Figure (\ref{fig:RP}) shows the plots of optimal trajectories in different reference levels $R$, assuming $x_0=1$ and $A=0$. Again, we end up with a unit TC order when $R=x_0$ and a unit IS order when $R=A$. 

As shown in Figure \ref{fig:RP}, when $R>x_0=1$ the optimal strategy suggest overshoot the target, meaning buying more than needed then sell back later and quicker when time approaches the terminal time. On the other hand, if $R < A = 0$, the optimal strategy suggests undershoots the target. We provide conditions on the parameters under which an overshooting or an undershooting occur in the following proposition.
\begin{proposition} \label{prop:over_shooting}
    Without loss of generality, assume $x_0 > A$. Let $x_t$ be the optimal strategy given in \eqref{eqn:opt-x-const}. We have that, if $R+\frac{\mu}{\theta \sigma^2}>x_0$, then $x_t$ is concave in $t$ and it becomes convex in $t$ if $R+\frac{\mu}{\theta \sigma^2}<A$. It follows that $x_t$ overshoots if and only if
        \[
            R+\frac{\mu}{\theta \sigma^2} > x_0 + \frac{1}{\cosh \kappa T - 1}\cdot (x_0-A)
        \]
        and it undershoots if and only if
        \[
           R+\frac{\mu}{\theta \sigma^2} < A - \frac{1}{\cosh \kappa T -1} \cdot (x_0-A).
        \]    
    \end{proposition}


It appears that, when $\mu$ is nonzero, the terms in the optimal trajectory \eqref{eqn:opt-x-const} that have $\mu$ involved always come in the form of $\frac{\mu}{\theta\sigma^2}$. We provide a financial rationale of this combination of parameters as follows. Note that the combination resembles the Merton ratio in Merton's optimal portfolio problem in the market consisting of two assets: one risky and the other risk-less. The agent is risk averse with power utility $u(x) = x^\gamma$ and the dynamic of risky asset is governed by a geometric Brownian motion with expected return $\mu$ and volatility $\sigma$.  The risk-less asset is assumed accruing zero interest. In this setting, the Merton ratio, which represents the percentage of wealth invested in the risky asset, is given by $\frac\mu{(1 - \gamma)\sigma^2}$. In our setting, the broker is with exponential utility $u(x)= \frac{1}{\theta}(1-e^{-\theta x})$ and the stock price $S_t$ is assumed following an arithmetic Brownian motion with drift $\mu$ and volatility $\sigma$, i.e.,  
\[ 
\ddd S_t = \mu \ddd t + \sigma \ddd W_t. 
\]
As in the derivation of the Merton ratio in Merton's problem, one may show that the ``Merton ratio'' \cite{merton1975optimum} in this case is given by $\frac{\mu}{\theta\sigma^2}$ except that this ratio does not represent the percentage of wealth invested in the risky asset as in the original Merton's problem. Indeed, this ratio represents the number of shares to be held in the optimal portfolio.
\subsection{Piece-wise constant reference strategy} \label{sec:5}
We consider in this section the reference strategies that are piece-wise constant across time. The consideration is that, to possibly account for real time market environments during the entire reallocation process, the broker's strategies may be subject to interval TWAP or VWAP strategies. For example, in an intraday trading activity, it is documented that the market trades much more actively at the times close to opening and closing than that in the middle of common trading days. Also, in the case where trading horizon across multiple days, the broker is most likely treating each trading day individually. In these regards, a piece-wise constant reference strategy is considered plausible. 

\begin{figure}[htbp]
    \centering
	\includegraphics[scale=0.5]{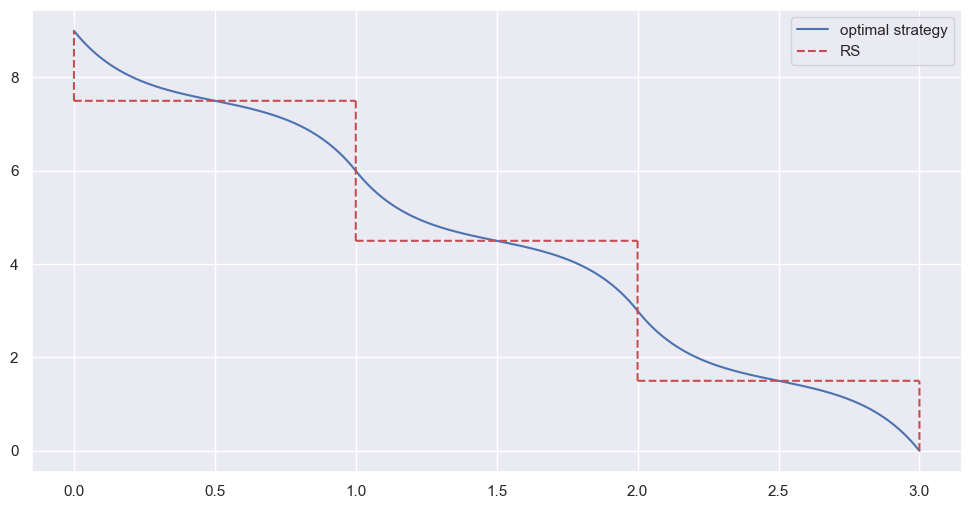}
	\caption{A three-period piece-wise constant reference strategy and its corresponding optimal trajectory.}
	\label{fig:threeperiod}
\end{figure}

Theoretically, as will be demonstrated, optimal trajectories subject to piece-wise constant reference strategies exhibit an elegant algebraic structure, derived from the unit IS and unit TC orders discussed in Sections \ref{subsec:4.2}. By applying standard approximation and limiting processes for integrable functions, this structure offers practical and insightful guidance for developing optimal strategies under general integrable reference strategies. Concretely, a generic piece-wise constant reference strategy $R_t$ is defined as
\begin{equation} \label{eqn:pwc-res-strat}
    R_t = (x_0 - R^{(1)}) \, \mathbb{1}_{\{t=0\}} + \sum_{k=1}^n R^{(k)} \, \mathbb{1}_{\left\{\frac{(k-1)T}{n} \le t < \frac{kT}n \right\}} + A \, \mathbb{1}_{\{t=T\}},
\end{equation}
where $\mathbb{1}_{\{\cdot\}}$ denotes the indicator function, $n$ is the number of periods, and the $R^{(k)}$'s are fixed constants. The following proposition summarizes the optimal trajectory under the piece-wise constant reference strategy in \eqref{eqn:pwc-res-strat}.
\begin{proposition} \label{prop:piecewise}
    When $\mu=m_0=0, \beta \to +\infty$, denote $R^{(0)}:=x_0$ and $R^{(n+1)}:=A$, the solution to the previous optimization problem under the piece-wise constant reference strategy given in \eqref{eqn:pwc-res-strat} is of the following form, for $1\le k\le n$,
    \begin{equation} \label{eqn:opt-x-pwc}
        x_t =  \left\{ a_k + \left( R^{(k)} - a_k \right) \, {\rm TC}_{t-\frac{(k-1)T}{n}} + \left(a_{k-1}-R^{(k)}\right) \, {\rm IS}_{t-\frac{(k-1)T}{n}}\right\} \, \mathbb{1}_{\left\{\frac{(k-1)T}{n} \le t < \frac{kT}{n} \right\}},
    \end{equation}
    with $\kappa:= \sqrt{\frac{\theta \sigma^2}{2\eta}}$, 
    \[
    {\rm IS}_t := \frac{\sinh \kappa \left(\frac{T}{n}-t\right)}{\sinh \kappa \frac{T}{n}}, \qquad {\rm TC}_t := \frac{\sinh \kappa \frac{T}{n}-\sinh \kappa t}{\sinh \kappa \frac{T}{n}},
    \] 
    and $a_k$ is denoted as a weighted average of $R^{(i)}$'s:
    \[
    a_k := \frac{\sinh \kappa\left(T-\frac{kT}{n}\right)}{\sinh \kappa T} \cdot \sum_{i=0}^k b_i R^{(i)}  + \frac{\sinh \kappa\frac{kT}{n}}{\sinh \kappa T} \cdot \sum_{i=k+1}^{n+1} b_{n-i+1} R^{(i)}, \ 1\le k\le n-1,
    \]
    where
    \[
    \begin{aligned}
    b_i =\begin{cases}
        1, & i=0,\\
        \cosh \left( \kappa \frac{iT}{n} \right) - \cosh \left( \kappa \frac{(i-1)T}{n} \right), & 1\le i\le n-1.
    \end{cases}
    \end{aligned}
    \]
\end{proposition}
Notice that the ${\rm IS}_t$ and ${\rm TC}_t$ in Proposition \ref{prop:piecewise} are indeed respectively the unit IS order given in \eqref{eq:IS} and the unit TC order in \eqref{eq:TC}. Proposition \ref{prop:piecewise} essentially states that, for a given piece-wise constant reference strategy, its corresponding optimal solution in each sub-interval is given by an affine transformation of a unit IS order and a unit TC order. In this sense we may regard the unit IS orders and the unit TC orders obtained in Sections \ref{subsec:4.2} as the building blocks or bases for optimal trajectories under piece-wise constant reference strategies. The algebraic structure of the optimal trajectories is thus depicted by the affine space generated by the unit IS and unit TC orders. What are left to be determined are the coefficients that are sub-interval dependent. In practice, one should firstly calculate $a_k$'s by linear combinations of $R^{(i)}$'s, then optimal strategy is obtained by connecting $a_k$'s using unit IS and unit TC orders. As an example, Figure (\ref{fig:threeperiod}) shows the plots of a piece-wise constant reference strategy (in dotted red) its optimal trajectory (in blue) with three sub-intervals $n=3$, the other parameters are given by $T=3$, $\kappa=5$, $R^{(1)}=\frac{15}{2}, R^{(2)}=\frac{9}{2}, R^{(3)}=\frac{3}{2}, x_0=9, A=0$.
To conclude the section, we show in Theorem \ref{thm:approx} a stability theorem for optimal strategies subject to general, smooth enough, reference strategies. 
 \begin{theorem} \label{thm:approx}
     For any given reference strategies $R_t$ and $\widetilde{R}_t$ satisfying
     \[
        \int_0^T \left(R_t-\widetilde{R}_t\right)^2 \ddd t < \varepsilon,
     \]
     for some $\varepsilon > 0$. Let $x_t$ and $\widetilde{x}_t$ be the corresponding optimal trajectories for $R_t$ and $\widetilde R_t$, respectively. We have that, for any $0\le t\le T$,
     \[
        \left| x_t-\widetilde{x}_t \right| < \frac{1}{2} \sqrt{\kappa \varepsilon}.
     \]
 \end{theorem}
The above theorem basically states the fact that when two reference strategies are close in the $L^2$ sense, their corresponding optimal strategies remain close in the sup norm sense. Note that for any $\varepsilon > 0$ and an almost everywhere continuous bounded RS $R_t$, there exists a piece-wise constant RS $\widetilde{R}_t$ such that
    \[ \int_0^T \left(R_t-\widetilde{R}_t\right)^2 \text{d}t < \varepsilon, \]
It follows that the optimal trajectory in \eqref{eqn:opt-x-pwc} for $\widetilde R_t$, which is an affine transformation of unit IS and unit TC orders in each sub-interval, can be applied as an approximation to the optimal trajectory under reference strategy $R_t$, with error estimate given in Theorem \ref{thm:approx}.
\section{Numerical examples} \label{sec:numerics}
We conduct in this section numerical experiments on the implementation of optimal strategy obtained in Theorem \ref{thm:sto_ctrl} and stress testing the strategy against various parameters. 
Monte Carlo simulations are implemented to illustrate sample trading trajectories and the performances criteria of the optimal strategy versus those of a related TWAP strategy.
The performance of the optimal and TWAP strategies are then stress tested against certain extreme parameters. 
We remark that the parameters chosen for the numerical examples in this section are for convenience only. In practice, parameters are supposedly calibrated to market data prior to implementation, causing possible issues associated with estimation risk.

\subsection{Sample trading trajectories}
\begin{figure}[htbp]
    \centering
	\includegraphics[scale=0.5]{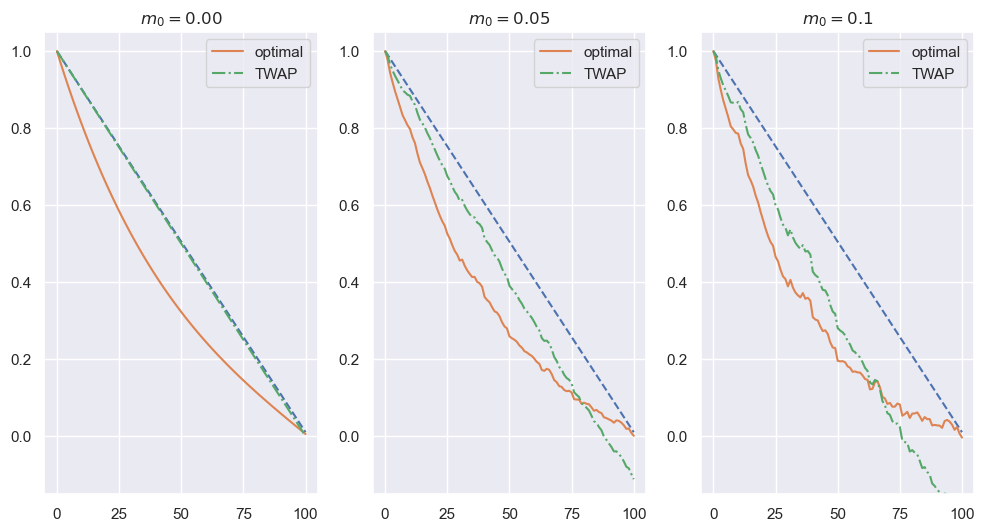}
	\caption{Sample trading trajectories in different $m_0$}
	\label{fig:m0}
    \end{figure}

\begin{figure}[htbp]
    \centering
	\includegraphics[scale=0.5]{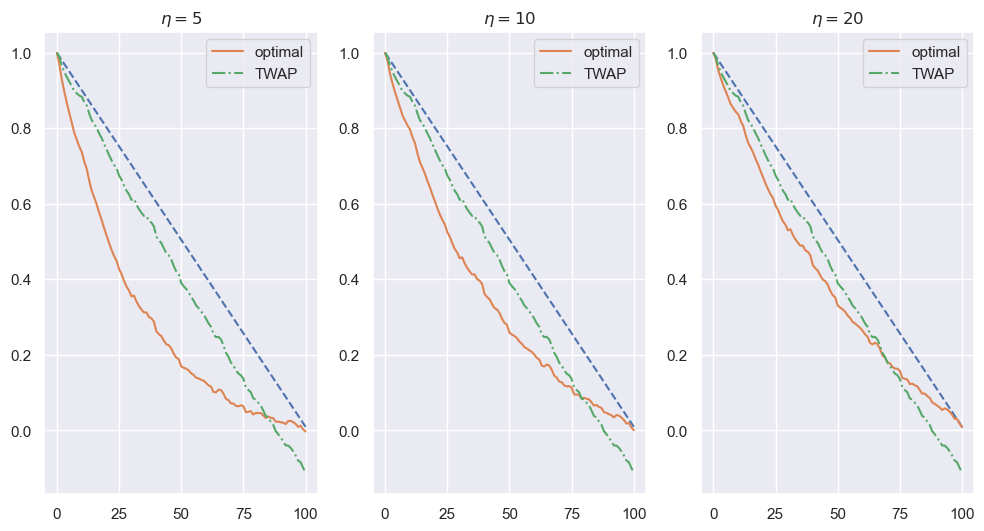}
	\caption{Sample trading trajectories in different $\eta$}
	\label{fig:eta}
    \end{figure}

\begin{figure}[htbp]
    \centering
	\includegraphics[scale=0.5]{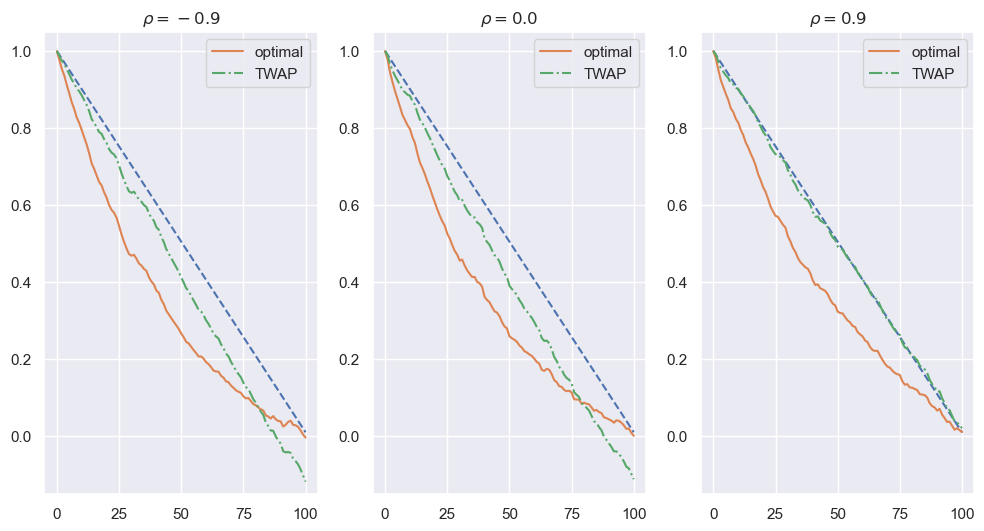}
	\caption{Sample trading trajectories in different $\rho$}
	\label{fig:rho}
    \end{figure}

\begin{figure}[htbp]
    \centering
	\includegraphics[scale=0.5]{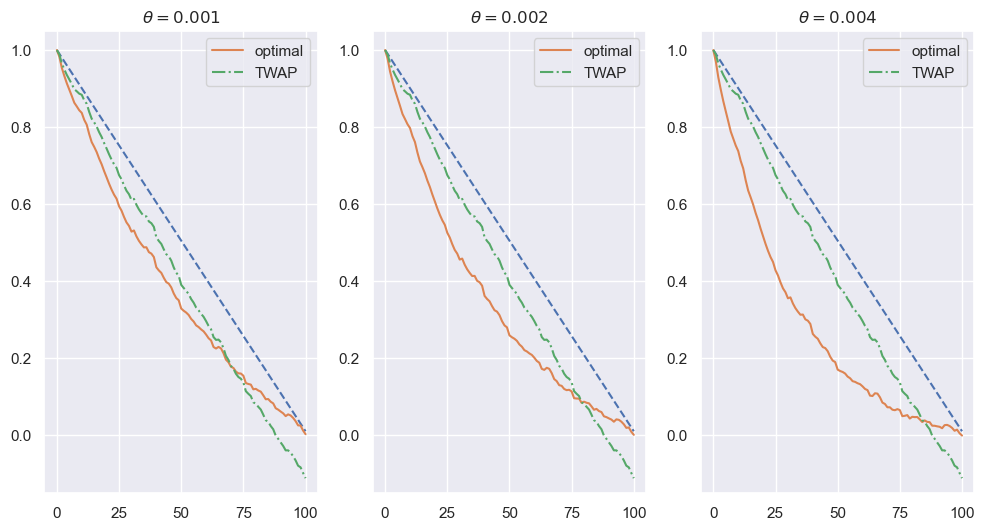}
	\caption{Sample trading trajectories in different $\theta$}
	\label{fig:theta}
    \end{figure}

\begin{figure}[htbp]
    \centering
	\includegraphics[scale=0.5]{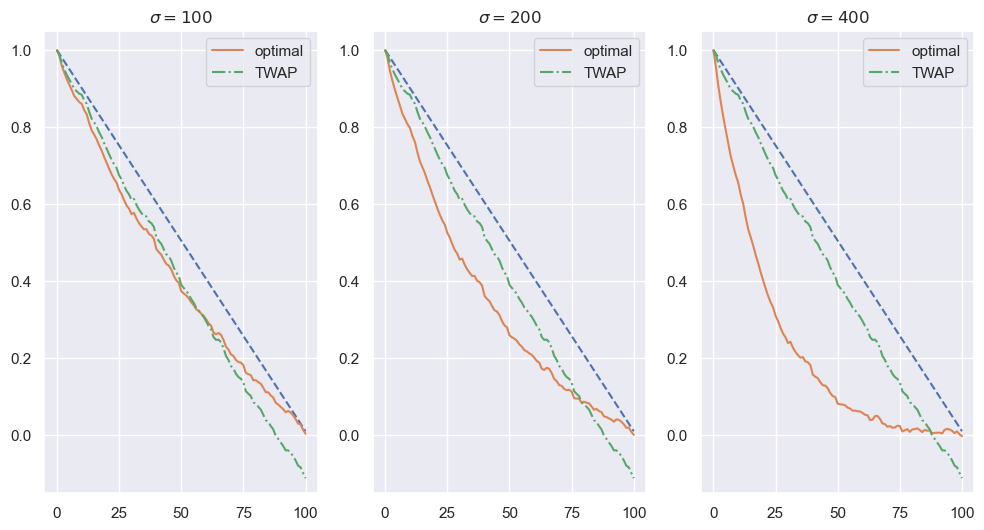}
	\caption{Sample trading trajectories in different $\sigma$}
	\label{fig:sigma}
    \end{figure}

\begin{figure}[htbp]
    \centering
	\includegraphics[scale=0.5]{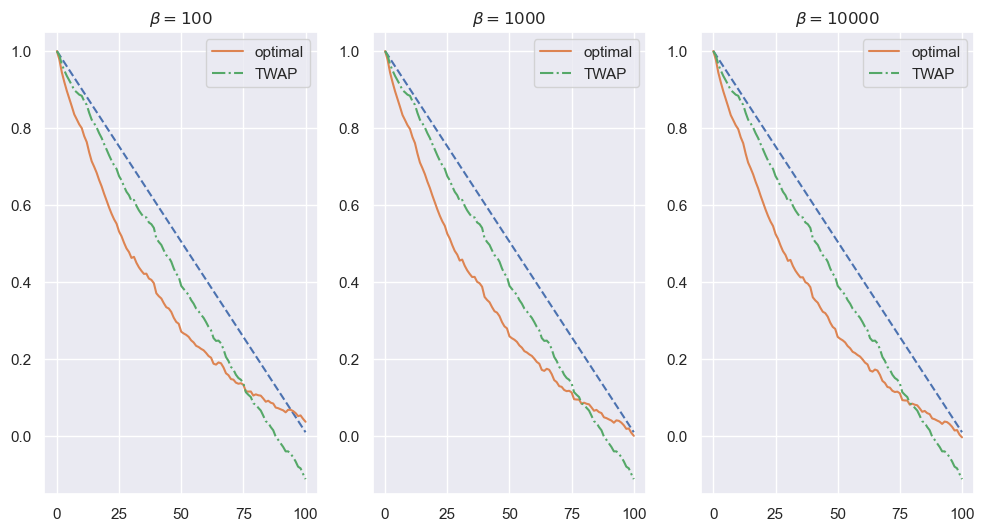}
	\caption{Sample trading trajectories in different $\beta$}
	\label{fig:beta}
    \end{figure}

We present in Figures (\ref{fig:m0}) through (\ref{fig:beta}) sample trading trajectories under the optimal and a related TWAP strategies in various parameters. Parameters chosen as base case are $(m_0, \eta, \rho, \theta, \sigma, \beta) = (0.05, 10, 0, 0.002, 200, 1000)$. 
Table \ref{tab:sample_traj} summarizes the parameters that vary, while the others are held fixed, in each case and their corresponding figures. The reference strategy $R_t$ is selected as a block trade at $t=0$, as shown in Section \ref{subsec:4.2}.

\begin{table}[]
\centering
\caption{Sample trading trajectory in varying parameters}
\label{tab:sample_traj}
\begin{tabular}{|c|c|c|c|c|}
\hline
Figure number    & Parameter & Case 1 & Case 2 & Case 3 \\ \hline \hline
Figure \ref{fig:m0} & $m_0$ & 0.00  & 0.05     & 0.10  \\ \hline
Figure \ref{fig:eta} & $\eta$ & 5  & 10     & 20  \\ \hline
Figure \ref{fig:rho} & $\rho$ & -0.9  & 0     & 0.9  \\ \hline
Figure \ref{fig:theta} & $\theta$ & 0.001     & 0.002     & 0.004  \\ \hline
Figure \ref{fig:sigma} & $\sigma$ & 100  & 200    & 400  \\ \hline
Figure \ref{fig:beta} & $\beta$ & 100  & 1000    & 10000  \\ \hline
\end{tabular}
\end{table}

Figure (\ref{fig:m0}) illustrates sample trading trajectories for the optimal and the TWAP strategies in different values of $m_0$. We note that, since $m_0$ quantifies the execution risk, as $m_0$ increases both trajectories become more volatile and fluctuating. 

    \newpage

%

Figure (\ref{fig:eta}) exhibits sample trading trajectories in varying values of $\eta$. It is worth mentioning that, since $\eta$ reflects the level of transaction costs, as the value of $\eta$ increases, transaction cost becomes more significant in the determination of optimal strategies. Consequently, the optimal strategy tends to align more closely with the classical TWAP strategy. In other words, a higher $\eta$ places greater emphasis on minimizing transaction costs, motivating the adoption of a strategy that mirrors the TWAP approach, which aims for consistent execution over time.

Figure (\ref{fig:rho}) shows optimal strategies in the values of $\rho=-0.5,0,0.5$. Recall that $\rho$ denotes the correlation between the stock price and the execution risk. We note that there seems no significant dependence of the optimal strategies on $\rho$ within this set of parameters.

%

\newpage

Figure (\ref{fig:theta}) illustrates the optimal strategies in the values of $\theta=0.001,0.002,0.004$, which represents the broker's taste of risk aversion. The larger the value of $\theta$, the more risk averse the broker tends to be. It follows that the sample trading trajectory under optimal strategy gradually converges towards the reference strategy as $\theta$ increases, indicating the broker's inclination to adopt a more conservative and risk-averse approach.
%
%

Figure (\ref{fig:sigma}) shows sample trading trajectories under optimal strategies with respect to the stock volatility $\sigma=100,200,400$. As $\sigma$ increases, the price risk increases. Thus, to mitigate the risk incurred from the price volatility during execution, the optimal strategy tracks more closely to the reference strategy.
%

\newpage

Figure (\ref{fig:beta}) presents the optimal strategies in the values of $\beta=100,1000,10000$, which quantifies the penalty of block trade at terminal time. Large $\beta$ indicates that any remaining inventory at terminal time is unfavorable. Therefore, it is seen in the figure that the terminal inventory $x_T^*$ of the sample trading trajectories under optimal strategy gradually approaches zero. It is thus fathomable that, as $\beta$ approaches infinity, a final block trade at terminal time turns into strictly prohibited. 
%



\subsection{Performance analysis and stress test}
To demonstrate the 
performance of the optimal strategy against TWAP, we implement the strategies using Monte Carlo simulations and present the resulting histograms and boxplots of the terminal wealths as well as the utilities at the investment horizon. 
This performance analysis serves to illustrate the optimality and robustness of the optimal strategy. 
\begin{figure}[htbp]
    \centering
	\includegraphics[scale=0.4]{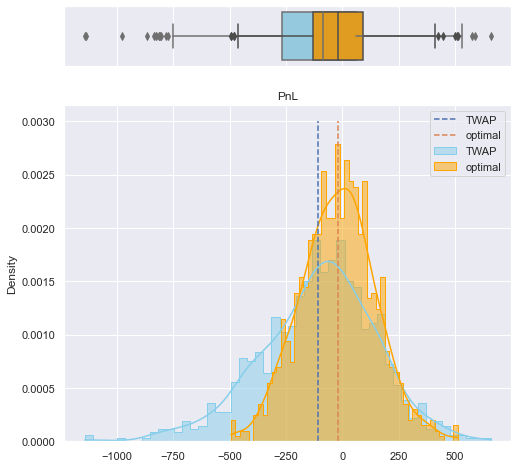} \quad 
	\includegraphics[scale=0.4]{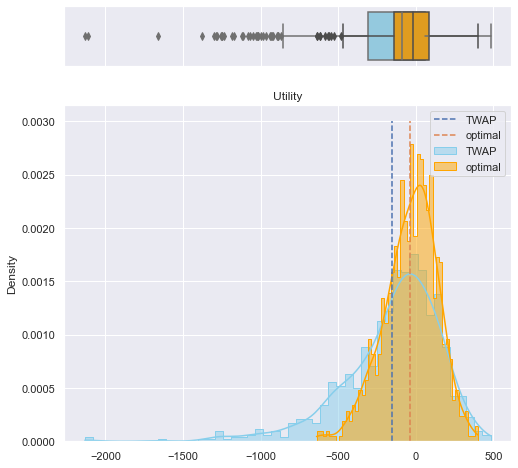}
	\caption{Histograms and boxplots of terminal wealths and utilities under the optimal and the TWAP strategies. Terminal wealth on the left, utility on the right. Vertical dashed lines indicate sample means.}
	\label{fig:PnL}
    \end{figure}

Figure (\ref{fig:PnL}) exhibits the histograms and boxplots of the terminal P\&L (on the left) and the utility (on the left) at investment horizon for the optimal and the TWAP strategies. We observe that the optimal strategy not only consistently achieves higher average terminal wealth and utility than those under TWAP but also bears lower variability. Furthermore, an intriguing observation is made regarding the tail ends of the distribution. Specifically, the TWAP strategy demonstrates a greater density in these tail regions, indicating a higher likelihood of incurring significant losses when compared to our optimal strategy. This discrepancy in density at the tail end reinforces the notion that the TWAP strategy may accentuate the potential for unfavorable outcomes, thus reflecting a subpar approach to risk management.

\begin{figure}[htbp]
    \centering
	\includegraphics[width=6cm, height=6cm]{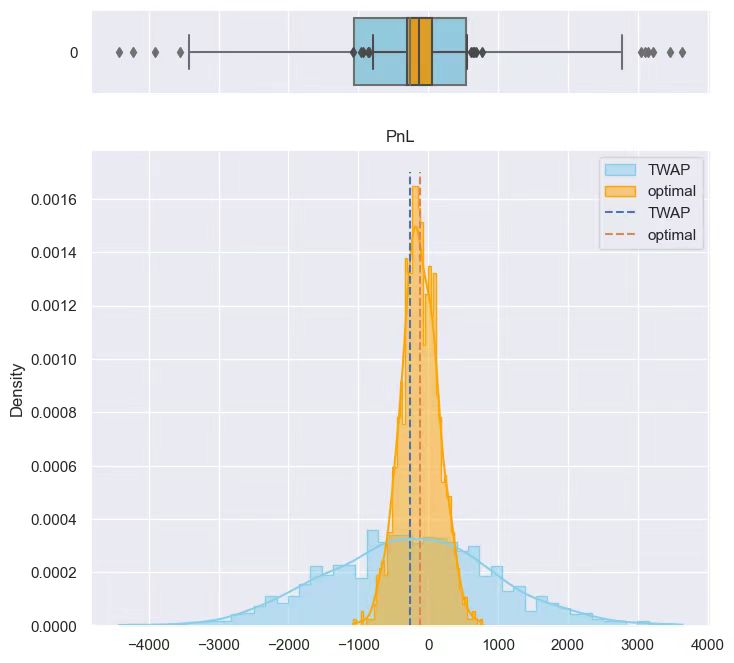} \qquad
	\includegraphics[width=6cm, height=6cm]{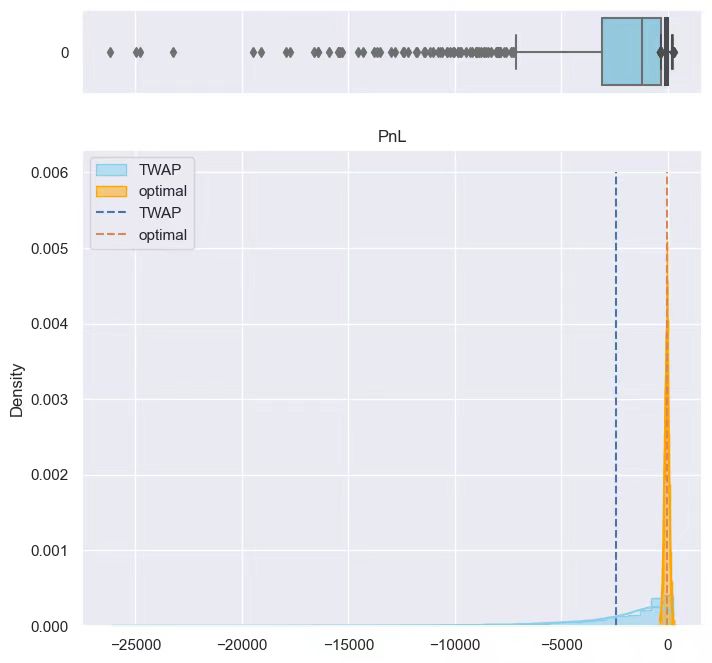} \\
	\vskip0.3cm
	\includegraphics[width=6cm, height=6cm]{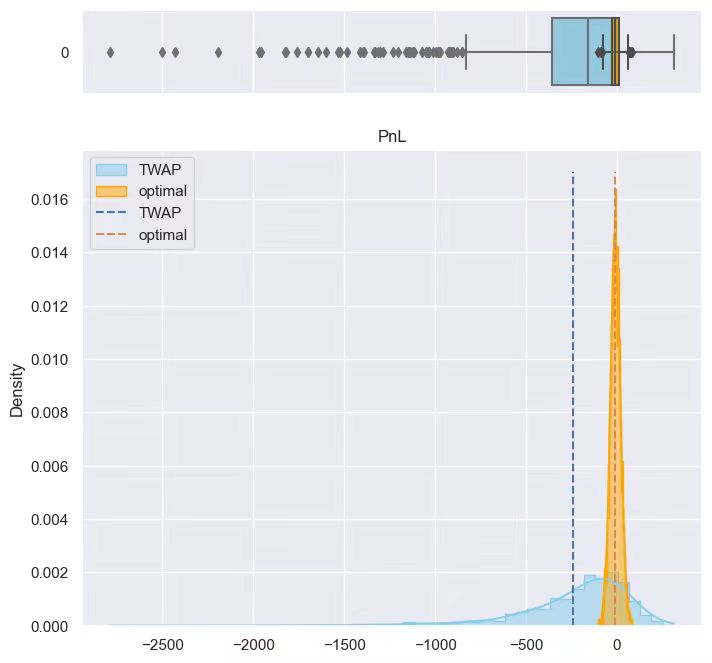} \qquad
	\includegraphics[width=6cm, height=6cm]{figure_stress_test/stress_test_eta.jpg}
	\caption{Histograms and boxplots of terminal P\&L under stress tests. Scenarios 1 on the top left, 2 on the top right, 3 on the bottom left, and 4 on the bottom right. Vertical dashed lines indicate sample means.}
	\label{fig:stress_test_sigma}
\end{figure}

Finally, we stress test the optimal and the TWAP strategies against extreme parameters. While the remaining parameters are held the same as in the benchmark case, the stock volatility $\sigma$ is scaled up by ten folds in Scenario 1, the terminal penalty coefficient $\beta$ up by ten folds in Scenario 2, the execution risk $m_0$ up by a hundred folds in Scenario 3, and lastly in Scenario 4 down by 0.1 hundred folds the temporary cost $\eta$. Table \ref{tab:stress_test} summarizes the parameters that are stress tested in each scenario and Figure \ref{fig:stress_test_sigma} shows the histograms and boxplots of P\&Ls at the terminal time under the optimal and the TWAP strategies.

Apparently, in all the scenarios the histograms under optimal strategies are concentrated around zero whereas those under TWAP strategies are much more widely spreading. Also, the sample means under optimal strategies are higher than those of TWAP strategies. We conclude that, even under extreme situations, the optimal strategies outperform TWAP strategies not only in higher averaged P\&L but also with lower volatility risk, which shows the robustness of our results.

%
\begin{table}[]
\centering
\caption{Stress testing parameters}
\label{tab:stress_test}
\begin{tabular}{|c|c|c|c|c|c|c|}
\hline
                          & $m_0$ & $\eta$ & $\rho$ & $\theta$ & $\sigma$ & $\beta$ \\ \hline
Baseline & 0.05  & 10     & 0      & 0.002    & 200     & 1e6  \\ \hline
Scenario 1 (large $\sigma$) & 0.05  & 10     & 0      & 0.002    & 2000     & 1e6  \\ \hline
Scenario 2 (large $\beta$)  & 0.05  & 10     & 0      & 0.002    & 200      & 1e7 \\ \hline
Scenario 3 (large $m_0$)    & 5     & 10     & 0      & 0.002    & 200      & 1e6  \\ \hline
Scenario 4 (small $\eta$) & 0.05  & 0.1    & 0      & 0.002    & 200      & 1e6  \\ \hline
\end{tabular}
\end{table}

\newpage

\section{Conclusion}
In this article, we showed how the broker's order execution problem under execution risk subject to client's reference strategy can be recast as a utility maximization problem. The optimal solution to the utility maximization problem was obtained by solving an associated Riccati differential equation and was presented in feedback form. When execution risk vanishes, the optimal strategies become deterministic and were given in closed form. We showed that trading trajectories subject to the unit IS and the unit TC orders form a basis of an affine structure for trading trajectories under optimal strategies subject to general piece-wise constant reference strategies. As for general continuous reference strategies, we proved an approximation and stability theorem for the trading trajectory under the corresponding optimal strategy via the trajectories from piece-wise constant reference strategies.

Numerical experiments showed that the optimal strategies not only achieve higher expected values but also bear lower variabilities as opposed to those under TWAP strategies. Numerical performance analysis and stress tests confirmed that optimal strategies outperform TWAP strategies, by a wide margin in certain cases. Possible extensions of the framework considered in the paper include (a) a stochastic component to the price evolution reflecting the overall market activity such macroeconomics factors or news; (b) the price impact, rather than permanent, being transient with proper decay kernel. We leave these considerations to a future study.

\section*{Acknowledgement}
This research is supported by the National Natural Science Foundation of China (Grants No.11971040).

\appendix
\section{Appendix} \label{sec:Appendix}

\addtocontents{toc}{\protect\setcounter{tocdepth}{0}}
\subsection{Proof to Proposition \ref{prop:PnL}} \label{proof:prop:PnL}

By It\^{o}'s formula,
\[
    \begin{aligned}
        (x_T-A)S_T =& (x_0-A)S_0 + \int_0^T S_t \dd x_t + \int_0^T \left( \mu(x_t-A) +\rho \sigma m(v_t) -\gamma v_t(x_t-A)  \right) \dd t \\
        &+ \int_0^T \sigma(x_t-A) \dd W_t + \int_0^T \gamma(x_t-A)m(v_t)\dd Z_t.
    \end{aligned}
    \]
    Furthermore, note that $\tilde{S}_t = S_t + \eta v_t$,
    \[
        \int_0^T\left(-\tilde{S}_t\right)\ddd x_t = -\int_0^T S_t \dd x_t - \int_0^T \eta v_t^2 \dd t + \int_0^T \eta v_t m(v_t)\dd Z_t,
    \]
    in summary,
    \[
    \begin{aligned}
        \Pi =& (x_T-A)S_T-\beta(x_T-A)^2+\int_0^T\left(-\tilde{S_t}\right)\ddd x_t \\
        =& (x_0-A)S_0-\beta(x_T-A)^2 +\int_0^T \left(\mu (x_t-A)+\rho \sigma m(v_t)-\eta v_t^2-\gamma v_t(x_t-A) \right) \ddd t \\
    &+ \int_0^T \sigma (x_t-A) \ddd W_t + \int_0^T (\eta v_t+\gamma (x_t-A))m(v_t)\ddd Z_t.
    \end{aligned}
    \]
    $\hfill \square$

\subsection{Risk-Neutral Case of Theorem \ref{thm:sto_ctrl}} \label{Appendix:risk-neutral}

In Section (\ref{sec:3}) we recast the optimal execution problem as a utility maximization problem with the CARA preference $u(x)=\frac{1}{\theta}(1-e^{-\theta x})$ with parameter $\theta>0$, which reflects the broker's risk aversion. As $\theta \to 0$, the utility function $u(x)=\lim_{\theta \to 0} \frac{1}{\theta}(1-e^{-\theta x})=x$ describes a risk-neutral preference. With the admissible set $\mathcal{A}$ and the excess P\&L $\tilde{\Pi}$ given in Section (\ref{sec:2.2}) and Section (\ref{sec:3}), the risk-neutral expected utility is given by
\begin{align*}
    \mathbb{E}\left[ u\left( \tilde{\Pi} \right) \right] =  \mathbb{E}\left[ \tilde{\Pi} \right] = & \mathbb{E} \Bigg[ -\beta (x_T-A)^2+\int_0^T \left(\mu (x_t-R_t)+\rho \sigma m(v_t)-\eta v_t^2-\gamma v_t(x_t-A) \right) \dd t\\
    &+ \int_0^T \sigma (x_t-R_t) \ddd W_t + \int_0^T (\eta v_t+\gamma (x_t-A))m(v_t)\dd Z_t \Bigg] \\
    =& \mathbb{E} \left[ -\beta (x_T-A)^2+\int_0^T \left(\mu (x_t-R_t)+\rho \sigma m(v_t)-\eta v_t^2-\gamma v_t(x_t-A) \right) \dd t \right].
\end{align*}
When the execution risk is constant (i.e. $m(\cdot)\equiv m_0$ for some $m_0\ge 0$), the utility maximization problem reduces to the following Stochastic Linear Quadratic (SLQ) control problem \cite{yong2012stochastic}:
\begin{align*}
    \begin{cases}
        \sup_{v \in \mathcal{A}} \mathbb{E} \left[ -\beta (x_T-A)^2+\int_0^T \left(\mu (x_t-R_t)+\rho \sigma m_0 -\eta v_t^2-\gamma v_t(x_t-A) \right) \dd t \right],\\
        \text{s.t. } x_t = x_0 - \int_0^t v_s \dd s + m_0 Z_t.
    \end{cases}
\end{align*}
Define the value function $V(t,x)$ as
\[
    V(t,x) = \sup_{v \in \mathcal{A}_t} \mathbb{E} \left[ -\beta (x_T-A)^2+\int_t^T \left(\mu (x_s-R_s)+\rho \sigma m_0 -\eta v_s^2-\gamma v_s(x_s-A) \right) \dd s \middle| x_t=x \right],
\]
where $\mathcal{A}_t:=\left\{ v \text{ is progressively measurable in } [t,T] \text{ and } \int_t^T \mathbb{E}[v_s^2] \dd s < \infty \right\}$. Note that our problem satisfies assumption \textbf{(L1)} given in Section 3.1 of \cite{yong2012stochastic} due to the boundedness of the coefficients, so the problem is solvable, and by the dynamic programming principle, the value function $V$ is unique satisfying the Hamilton-Jacobi-Bellman (HJB hereafter) equation
\begin{align*}
    V_t + \frac{1}{2}m_0^2 V_{xx} + \sup_{v\in \mathbb{R}} \left\{-\eta v^2- v\left(\gamma(x-A)+V_x\right) \right\} + \mu(x-R_t)+\rho \sigma m_0 = 0
\end{align*}
with terminal condition $v(T,x)=-\beta(x-A)^2$. With first order criterion, the optimal feedback control is given by $v^* = -\frac{1}{2\eta}\left(\gamma(x-A)+V_x\right)$, hence the HJB equation reduces to 
\begin{align*}
    V_t + \frac{1}{2}m_0^2 V_{xx} +\frac{1}{4\eta}\left(\gamma(x-A)+V_x\right)^2 + \mu(x-R_t)+\rho \sigma m_0 = 0, \ \ v(T,x)=-\beta(x-A)^2.
\end{align*}
Assume the Ansarz for value function $V$
\[
    V(t,x) = a(t)x^2 + b(t)x + c(t).
\]
Plug the Ansarz into the HJB equation yields the Riccati ODE system
\begin{align*}
\begin{cases}
    a'(t)+\frac{1}{4\eta}(2a(t)+\gamma)^2 = 0, & a(T)=-\beta,\\
    b'(t)+\frac{1}{2\eta}(2a(t)+\gamma)(b(t)-\gamma A)+\mu = 0, & b(T)=2\beta A,\\
    c'(t)+m_0^2 a(t)+\frac{1}{4\eta}(b(t)-\gamma A)^2-\mu R_t + \rho \sigma m_0 =0, & c(T)=-\beta A^2.
\end{cases}
\end{align*}
By solving the ODE system, the deterministic functions $a,b,c$ of time $t$ are given by
\begin{align*}
    a(t)=& -\frac{\eta}{T-t+\alpha}-\frac{\gamma}{2}, \\
    b(t)=& \frac{\mu}{2}(T-t+\alpha)-\frac{\frac{1}{2}\mu \alpha^2-2A\eta}{T-t+\alpha}+\gamma A,\\
    c(t)=& -m_0^2\eta \log \frac{T-t+\alpha}{\alpha} + \left(\rho \sigma m_0 - \frac{\gamma}{2}m_0^2\right)(T-t) \\ &- \frac{\mu^2\alpha^3}{16\eta} \left( \frac{\alpha}{T-t+\alpha} +\frac{2(T-t+\alpha)}{\alpha} - \frac{(T-t+\alpha)^3}{3\alpha^3} - \frac{8}{3} \right) \\
    &-\frac{A^2\eta-\frac{1}{2}\mu A\alpha^2}{T-t+\alpha}-\frac{\gamma A^2+\mu A(T-t+\alpha)}{2} + \int_t^T \mu(A-R_s)\dd s
\end{align*}
where $\alpha:=\frac{2\eta}{2\beta-\gamma}>0$. Note that $V(t,x)=a(t)x^2+b(t)x+c(t)$ is continuous on $[0,T]\times \mathbb{R}$ and continuously differentiable on $(0,T)\times \mathbb{R}$ satisfying the HJB equation, therefore it must be the value function by the uniqueness of solutions to the HJB equation. Then the verification theorem gives rise to the optimal feedback control
\begin{align*}
    v^{**}_t =& \frac{x_t-A}{T-t+\alpha} - \frac{\mu}{4\eta}\left( (T-t+\alpha) - \frac{\alpha^2}{T-t+\alpha} \right).
\end{align*}
Due to the boundedness of $\frac{1}{T-t+\alpha}$ and $-\frac{A}{T-t+\alpha}-\frac{\mu}{4\eta\left( (T-t+\alpha) - \frac{\alpha^2}{T-t+\alpha} \right)}$, $v^{**} \in \mathcal{A}$, which implies that the optimal control $v^{**}$ lies in the admissible set. Moreover, the maximal risk-neutral expected utility is given by
\begin{align*}
    \sup_{v\in \mathcal{A}} \mathbb{E}\left[ \tilde{\Pi} \right] =& \mathbb{E}\left[ \tilde{\Pi} \right] \bigg|_{v=v^{**}} = V(0,x_0)=a(0)x_0^2 + b(0)x_0 + c(0) \\
    =& \left(-\frac{\eta}{T+\alpha}-\frac{\gamma}{2}\right)x_0^2 + \left(\frac{\mu}{2}(T+\alpha)-\frac{\frac{1}{2}\mu \alpha^2-2A\eta}{T+\alpha}+\gamma A\right) x_0 \\
    & -m_0^2\eta \log \frac{T+\alpha}{\alpha} + \left(\rho \sigma m_0 - \frac{\gamma}{2}m_0^2\right)T  - \frac{\mu^2\alpha^3}{16\eta} \left( \frac{\alpha}{T+\alpha} +\frac{2(T+\alpha)}{\alpha} - \frac{(T+\alpha)^3}{3\alpha^3} - \frac{8}{3} \right) \\
    &-\frac{A^2\eta-\frac{1}{2}\mu A\alpha^2}{T+\alpha}-\frac{\gamma A^2+\mu A(T+\alpha)}{2} + \int_0^T \mu(A-R_s)\dd s < \infty.
\end{align*}

\subsection{Proof to Theorem \ref{thm:sto_ctrl}} \label{proof:thm:sto_ctrl}

Note that $x_t = x_0 - \int_0^t v_s\dd s + m_0 \dd Z_t$, by It\^{o}'s formula, 
\[
    (x_T-A)^2 = (x_t-A)^2 + \int_t^T \left( -2(x_s-A)v_s+m_0^2 \right) \dd s + \int_t^T \left( 2m_0(x_s-A) \right) \dd Z_s.
\]
Furthermore, note that standard Brownian motions $W_t$ and $Z_t$ have constant correlation $\rho$, denote standard Brownian motion $W_t^1 = \frac{Z_t-\rho W_t}{\sqrt{1-\rho^2}}$, then $W_t^1$ and $W_t$ are independent. By extracting the constants and deterministic parts out of the expectation, the value function is given by
\[
         \begin{aligned} 
              V(t,x) = \frac{1}{\theta} - \frac{1}{\theta} \cdot \exp \left( -\int_t^T \theta \left( \mu(A-R_s) + \rho \sigma m_0-\beta m_0^2 \right) \dd s + \theta \beta (x-A)^2 \right) \cdot V_0(t,x), 
              \end{aligned}
\]
where we observe the equivalent optimization problem
\[
\begin{aligned}
        V_0(t,x) =& \begin{aligned}[t]
                \inf_{v \in \mathcal{A}_t}  \mathbb{E}^{\mathbb{P}} \Bigg[ \exp\Bigg\{ &\int_t^T \left[ -\theta \mu (x_s-A) +\theta\eta v_s^2-\theta (2\beta-\gamma) v_s(x_s-A) \right] \dd s \\
              &+ \int_t^T \left[ \left(-\theta\sigma-\theta(\gamma-2\beta)m_0\rho\right) (x_s-A)-\theta \eta m_0\rho v_s+\theta \sigma (R_s-A) \right] \dd W_s \\
              &+ \int_t^T \left[-\theta (\gamma-2\beta) m_0\sqrt{1-\rho^2} (x_s-A)  -\theta \eta m_0 \sqrt{1-\rho^2} v_s\right]\dd W_s^1 \Bigg\} \Bigg | x_t=x  \Bigg].
              \end{aligned}\\
              =: & \inf_{v \in \mathcal{A}_t} \mathbb{E}^{\mathbb{P}} \left[ \exp \left\{ \textcolor{red}{J(t,T)} \right\} | x_t=x \right]
\end{aligned}
\]
Our stochastic control problem has a similar form as stochastic linear 
exponential quadratic (SLEQ) control problems given in \cite{duncan2013linear} and \cite{lim2005new}, however, due to the constants and stochastic integrals appearing in our exponential function, these existing results can't be used directly. Instead, we follow the ``completing the square'' technique used in \cite{duncan2013linear}, which is used to transform the integrand into a square form by introducing a Radon-Nikodym derivative. In this paper, we eliminate the stochastic integrals and ``complete the square'' simultaneously by Girsanov's theorem. More precisely, we manage to find some probability measure $\mathbb{Q}\ll \mathbb{P}$ such that
\[
    \mathbb{E}^{\mathbb{P}} \left[ \exp \left\{ J(t,T) \right\} |x_t=x \right] = \mathbb{E}^{\mathbb{Q}} \left[ \frac{1}{L(t,T)} \cdot \exp \left\{J(t,T) \right\} \bigg| x_t=x \right],
\]
where $\frac{1}{L(t,T)} \cdot \exp \left\{J(t,T) \right\}$ can be transformed into a square form without It\^{o} integrals. Concretely, suppose the Radon-Nikodym derivative is given by 
\[
    \begin{aligned}
        \frac{\dd \mathbb{Q}}{\dd \mathbb{P}}\bigg|_{\mathcal{F}_t} = \textcolor{blue}{L(t,T)}:= \exp\Bigg\{ -&\int_t^T \Psi_1(s,x_s,v_s)\dd W_s-\int_t^T \Psi_2(s,x_s,v_s)\dd W_s^1  \\
         -&\frac{1}{2} \int_t^T \left( \Psi_1^2(s,x_s,v_s)+\Psi_2^2(s,x_s,v_s) \right) \dd s \Bigg\},
    \end{aligned}
\]
where $\Psi_1$ and $\Psi_2$ are $L^2$-functions to be determined. At the same time, for some deterministic function $G_1(t)$ and $G_2(t)$ such that $G_1(T)=G_2(T)=0$, by Ito's lemma,
\[
\begin{aligned}
   &  \textcolor{cyan}{- G_1(t)(x_t-A)^2 - 2G_2(t)(x_t-A)} \\
   =& (G_1(T)(x_T-A)^2+2G_2(T)(x_T-A)) - (G_1(t)(x_t-A)^2 + 2G_2(t)(x_t-A)) \\
   =& \int_t^T \left( G_1'(s)(x_s-A)^2 +2G_2'(s)(x_s-A)-2(G_1(s)(x_s-A)+G_2(s))v_s + m_0^2G_1(s) \right) \dd s \\
   &+ \int_t^T 2\rho m_0(G_1(s)(x_s-A)+G_2(s))\dd W_s + \int_t^T 2\sqrt{1-\rho^2}m_0(G_1(s)(x_s-A)+G_2(s))\dd W_s^1.
\end{aligned}
\]
Hence,
\[
\begin{aligned}
    & \frac{1}{L(t,T)} \cdot \exp \left\{J(t,T) \right\} \\
    =& \exp \big\{ \textcolor{red}{J(t,T)}-\textcolor{blue}{\log L(t,T)}  +\textcolor{cyan}{\left( - G_1(t)(x_t-A)^2 - 2G_2(t)(x_t-A) \right)} + \left(G_1(t)(x_t-A)^2 + 2G_2(t)(x_t-A) \right) \big\} \\
    =& \begin{aligned}[t] \exp\Bigg\{ &\int_t^T \bigg[ \begin{aligned}[t] & \textcolor{red}{-\theta \mu (x_s-A) +\theta\eta v_s^2-\theta (2\beta-\gamma) v_s(x_s-A)} +\textcolor{blue}{ \frac{1}{2} \Psi_1^2(s,x_s,v_s)+ \frac{1}{2} \Psi_2^2(s,x_s,v_s)} \\
    &+ \textcolor{cyan}{\left( G_1'(s)(x_s-A)^2 +2G_2'(s)(x_s-A)-2(G_1(s)(x_s-A)+G_2(s))v_s + m_0^2G_1(s) \right)} \bigg] \dd s \end{aligned} \\
              &+ \int_t^T  \bigg[ \begin{aligned}[t] & \textcolor{red}{\left(-\theta\sigma-\theta(\gamma-2\beta)m_0\rho\right) (x_s-A)-\theta \eta m_0\rho v_s+\theta \sigma (R_s-A)} +\textcolor{blue}{ \Psi_1(s,x_s,v_s)} \\
                   & +\textcolor{cyan}{ 2\rho m_0(G_1(s)(x_s-A)+G_2(s))} \bigg]  \dd W_s \end{aligned} \\
                  &+ \int_t^T \bigg[ \begin{aligned}[t] & \textcolor{red}{-\theta (\gamma-2\beta) m_0\sqrt{1-\rho^2} (x_s-A)  -\theta \eta m_0 \sqrt{1-\rho^2} v_s} + \textcolor{blue}{ \Psi_2(s,x_s,v_s)} \\
                  & +\textcolor{cyan}{2\sqrt{1-\rho^2}m_0(G_1(s)(x_s-A)+G_2(s))} \bigg]\dd W_s^1 \end{aligned}\\
              &+  \left(G_1(t)(x_t-A)^2 + 2G_2(t)(x_t-A) \right) \Bigg\} \end{aligned} \\
\end{aligned}
\]
By setting
\[
\begin{cases}
\begin{aligned}
    \textcolor{blue}{\Psi_1(s,x_s,v_s)} =& ( \textcolor{red}{\theta\sigma}+m_0\rho(\textcolor{red}{\theta(\gamma-2\beta)}-\textcolor{cyan}{2G_1(s)}))&(x_s-A) &+& \textcolor{red}{\theta \eta m_0 \rho v_s} &- \textcolor{cyan}{2m_0\rho G_2(s)} - \textcolor{red}{\theta \sigma(R_s-A)}, \\
    \textcolor{blue}{\Psi_2(s,x_s,v_s)} =& \left( \textcolor{red}{\theta(\gamma-2\beta)}-\textcolor{cyan}{2G_1(s)} \right)m_0\sqrt{1-\rho^2}&(x_s-A) &+& \textcolor{red}{\theta \eta m_0\sqrt{1-\rho^2}v_s} &-\textcolor{cyan}{2m_0\sqrt{1-\rho^2}G_2(s)},
\end{aligned}
\end{cases}
\]
the stochastic integrals $\int_t^T \cdots \dd W_s$ and $\int_t^T \cdots \dd W_s^1$ appear to be zero. Moreover, 
\begin{equation} \label{eq:integrand}
\begin{aligned}
    & \frac{1}{L(t,T)} \cdot \exp \left\{J(t,T) \right\} \\
    =& \exp \Bigg\{ \begin{aligned}[t] & \int_t^T \left(  A(s)v_s^2 + B(s)(x_s-A)v_s + C(s)(x_s-A)^2 + D(s)v_s + E(s) (x_s-A) + F(s) \right) \dd t  \\
    & +  \left(G_1(t)(x_t-A)^2 + 2G_2(t)(x_t-A) \right) \Bigg\} \end{aligned} ,
\end{aligned}
\end{equation}
where the coefficients are given by
\[
    \begin{cases}
    \begin{aligned}
        H \equiv A(s) :=& \theta\eta + \frac{1}{2}\theta^2 \eta^2 m_0^2 > 0, \\
        B(s) =& -\theta(2\beta-\gamma)-2G_1(s)+ m_0\eta\theta^2 \rho \sigma + \theta \eta m_0^2(\theta(\gamma-2\beta)-2G_1(s)),\\
        C(s) =& G_1'(s) + \frac{1}{2}m_0^2\left( \theta(\gamma-2\beta)-2G_1(s) \right)^2 + \frac{1}{2}\theta^2\sigma^2 + \theta \sigma \rho m_0 \left( \theta(\gamma-2\beta)-2G_1(s) \right),\\
        D(s) =& -2G_2(s)-2m_0^2\theta \eta G_2(s)-  m_0\eta\theta^2 \rho \sigma(R_s-A),\\
        E(s) =& -\theta \mu + 2G_2'(s)-2m_0^2(\theta(\gamma-2\beta)-2G_1(s))G_2(s)-\theta^2\sigma^2(R_s-A)-2m_0\rho\theta \sigma G_2(s)\\
        &- m_0\rho \theta \sigma \left( \theta(\gamma-2\beta)-2G_1(s) \right)(R_s-A)\\
        F(s) =& m_0^2 G_1(s)+ 2m_0^2G_2^2(s)+2\rho m_0\theta \sigma(R_s-A)G_2(s) +\frac{1}{2}\theta^2 \sigma^2 (R_s-A)^2.\\
    \end{aligned}
    \end{cases}
\]
To make sure the integrand in (\ref{eq:integrand}) behaves as a square form (plus constants), we rewrite it as a quadratic function of $v_s$:
\[
    A(s) v_s^2 + \left( B(s)(x_s-A)+D(s) \right) v_s + \left( C(s)(x_s-A)^2+E(s)(x_s-A)+F_0(s) \right) + \left( F(s)-F_0(s) \right),
\]
of which the discriminant is set to be zero:
\[
    0 = \Delta = \left( B(s)(x_s-A)+D(s) \right)^2 - 4A(s) \left( C(s)(x_s-A)^2+E(s)(x_s-A)+F_0(s) \right),
\]
hence,
\[
    \begin{cases}
        \begin{aligned}
            0 =& B^2(s) - 4A(s)C(s),\\
            0 =& B(s)D(s) - 2A(s)E(s),\\
            0 =& D^2(s)-4A(s)F_0(s) \implies F_0(s) = \frac{D^2(s)}{4A(s)}.
        \end{aligned}
    \end{cases}
\]
Denote $b_2(s):=G_1(s)+\frac{\theta}{2}(2\beta-\gamma)$ and $b_1(s):=2G_2(s)$, we may solve the ODEs $b_2(\cdot)$ and $b_1(\cdot)$ satisfy:
$$
    b_2' = \frac{1}{H}((b_2+l_3)^2-l_1), \ \ b_2(T)=\frac{\theta}{2}(2\beta-\gamma),
$$
where $ l_3=\frac{1}{2}m_0\eta \theta^2 \rho \sigma$ and $l_1=\frac{\theta^2 \sigma^2}{2}H$. We may solve that
$$
    b_2(t) = \sqrt{l_1}\coth \left( A_0 + \frac{\sqrt{l_1}}{H}(T-t) \right)-l_3,
$$
where
$$
    A_0 := \coth^{-1} \left( \frac{l_3+\frac{\theta}{2}(2\beta-\gamma)}{\sqrt{l_1}} \right).
$$
Moreover, we have $b_1$ solves the ODE
$$
    b_1' = \theta \mu + \frac{1}{H}b_1(b_2+l_3) + \frac{\eta \theta^2\sigma}{H}(R_t-A)\left( \theta \sigma + \frac{1}{2}m_0^2\theta^2\eta \sigma (1-\rho^2)-\rho m_0 b_2 \right), \ \ b_1(T)=0.
$$
With $b_1$ and $b_2$ solved, we may rewrite the integrand in (\ref{eq:integrand}) as
\[
\begin{aligned}
    & A(s)v_s^2 + B(s)(x_s-A)v_s + C(s)(x_s-A)^2 + D(s)v_s + E(s) (x_s-A) + F(s) \\
    =& H \left( v_s - g_1(s)(x_s-A) - g_0(s) \right)^2 + \left( F(s)-\frac{D^2(s)}{4H} \right)\\
    =& H \left( v_s - g_1(s)(x_s-A) - g_0(s) \right)^2 + \left( m_0^2\left( b_2-\frac{\theta}{2}(2\beta-\gamma) \right) + \frac{\theta^2\sigma^2}{2}(R_s-A)^2-\frac{1}{4H}\left( b_1-\rho m_0\theta^2\sigma \eta (R_s-A) \right)^2 \right),
\end{aligned}
\]
where
$$
\begin{aligned}
    g_1(t) := \frac{(1+m_0^2\theta \eta)b_2-l_3}{H}, \ \ g_0(t):=\frac{(1+\theta \eta m_0^2)b_1+2l_3(R_u-A)}{2H}.
\end{aligned}
$$
Denote
\[
    c(s) := m_0^2\left( b_2-\frac{\theta}{2}(2\beta-\gamma) \right) + \frac{\theta^2\sigma^2}{2}(R_s-A)^2-\frac{1}{4H}\left( b_1-\rho m_0\theta^2\sigma \eta (R_s-A) \right)^2, \ 0\le  s\le T,
\]
we have
\[
\begin{aligned}
    & \mathbb{E}^{\mathbb{P}} \left[ \exp \left\{ J(t,T) \right\} | x_t=x \right] = \mathbb{E}^{\mathbb{Q}} \left[ \frac{1}{L(t,T)} \cdot \exp \left\{J(t,T) \right\} \bigg| x_t=x \right] \\
    =& \mathbb{E}^{\mathbb{Q}} \Bigg[ \begin{aligned}[t] & \exp \Bigg\{ \int_t^T H(v_s-g_1(s)(x_s-A)-g_0(s))^2 \dd s \Bigg\} \\
    & \cdot \exp \Bigg\{\left(b_2(t)-\frac{\theta}{2}(2\beta-\gamma)\right)(x_t-A)^2+b_1(t)(x_t-A) + \int_t^T c(s) \dd s \Bigg \} \Bigg| x_t=x \Bigg] \end{aligned} \\
    =& \mathbb{E}^{\mathbb{Q}} \Bigg[ \begin{aligned}[t] & \exp \Bigg\{ \int_t^T H(v_s-g_1(s)(x_s-A)-g_0(s))^2 \dd s \Bigg\} \Bigg| x_t=x \Bigg] \\
    & \cdot \exp \Bigg\{\left(b_2(t)-\frac{\theta}{2}(2\beta-\gamma)\right)(x-A)^2+b_1(t)(x-A) + \int_t^T c(s) \dd s \Bigg \}  \end{aligned} \\
    \ge & \exp \Bigg\{\left(b_2(t)-\frac{\theta}{2}(2\beta-\gamma)\right)(x-A)^2+b_1(t)(x-A) + \int_t^T c(s) \dd s \Bigg \},
\end{aligned}
\]
where the equation holds when $v_s = v_s^* := g_1(s)(x_s-A)+g_0(s)$. Note that $g_1$ and $g_0$ are bounded and continuous on $[0,T]$, so $v^* \in \mathcal{A}_t$, hence,
\[
\begin{aligned}
    V_0(t,x) =& \inf_{v\in \mathcal{A}_t} \mathbb{E}^{\mathbb{P}} \left[ \exp \left\{ J(t,T) \right\} |x_t=x \right] = \inf_{v\in \mathcal{A}_t} \mathbb{E}^{\mathbb{Q}} \left[ \frac{1}{L(t,T)} \cdot \exp \left\{J(t,T) \right\} \bigg| x_t=x \right] \\
    =& \mathbb{E}^{\mathbb{Q}} \left[ \frac{1}{L(t,T)} \cdot \exp \left\{J(t,T) \right\} \bigg| x_t=x \right]\bigg|_{v=v^*} \\
    =&  \exp \Bigg\{\left(b_2(t)-\frac{\theta}{2}(2\beta-\gamma)\right)(x-A)^2+b_1(t)(x-A) + \int_t^T c(s) \dd s \Bigg \}.
\end{aligned}
\]
Furthermore, the value function is given by
\[
         \begin{aligned} 
              V(t,x) =& \frac{1}{\theta} - \frac{1}{\theta} \cdot \exp \left\{ -\int_t^T \theta \left( \mu(A-R_s) + \rho \sigma m_0-\beta m_0^2 \right) \dd s + \theta \beta (x-A)^2 \right\} \cdot V_0(t,x) \\
              =& \frac{1}{\theta} - \frac{1}{\theta} \cdot \exp \Bigg\{ \begin{aligned}[t] & \left( b_2(t)+\frac{\theta\gamma}{2} \right) (x-A)^2 + b_1(t)(x-A) \\
              & +  \int_t^T \left[ c(s) -\theta \mu(A-R_s) - \theta\rho \sigma m_0+\theta\beta m_0^2  \right] \dd s \Bigg\}. \end{aligned} \\
              =& \frac{1}{\theta} \left( 1 - \exp \left\{ \left( b_2(t)+\frac{\theta\gamma}{2} \right) (x-A)^2 + b_1(t)(x-A) + b_0(t) \right\} \right),
              \end{aligned}
\]
where
\[
\begin{aligned}
    b_0(t):=& \int_t^T \left[ c(s) -\theta \mu(A-R_s) - \theta\rho \sigma m_0+\theta\beta m_0^2  \right] \dd s \\
    =& \int_t^T \left[ \frac{l_1-l_3^2}{H} (R_s-A)^2 + \left( \theta \mu + \frac{l_3}{H}b_1 \right)(R_s-A) + \left( -\frac{b_1^2}{4H}+m_0^2\left( b_2+\frac{\theta \gamma}{2} \right)-\rho m_0\theta \sigma \right) \right] \dd s.
\end{aligned}
\]
In summary, the optimal feedback control is given by
\[
\begin{aligned}
    v_t^* =&  g_1(t)(x_t-A)+g_0(t)\\
    =& \frac{(1+m_0^2\theta \eta)b_2-l_3}{H} \cdot (x_t-A) + \frac{l_3}{H}\cdot (R_t-A) + \frac{(1+m_0^2\theta \eta)b_1}{2H}.
\end{aligned}
\]
$\hfill \square$

\subsection{Proof to Theorem \ref{thm:det_ctrl}} \label{proof:thm:det_ctrl}

When $m(v)=m_0=0$ and $\beta=0$, $\dd x_t=-v_t\dd t$, according to Theorem \ref{thm:sto_ctrl}, $H=\theta \eta$, $l_3=0$, $l_1=\frac{\theta^3\sigma^2 \eta}{2}$, $A_0=0$, and
\[
    b_2(t)=\sqrt{\frac{\theta^3\sigma^2 \eta}{2}} \cdot \coth \left( \sqrt{\frac{\theta\sigma^2}{2\eta}}(T-t) \right).
\]
Furthermore, $b_1$ solves the backward ODE
\[
    b_1' = \theta \mu + \frac{1}{H}b_1b_2 +\theta^2 \sigma^2(R_t-A), \ \ b_1(T)=0.
\]
Note that
\[
\begin{aligned}
    \left(b_1 \cdot \sinh \left( \sqrt{\frac{\theta\sigma^2}{2\eta}}(T-t) \right)\right)'=& \sinh \left( \sqrt{\frac{\theta\sigma^2}{2\eta}}(T-t) \right) \cdot b_1' - b_1 \cdot \sqrt{\frac{\theta\sigma^2}{2\eta}} \cdot \cosh \left( \sqrt{\frac{\theta\sigma^2}{2\eta}}(T-t) \right) \\
    =&  \sinh \left( \sqrt{\frac{\theta\sigma^2}{2\eta}}(T-t) \right) \cdot \left( \theta \mu + \theta^2 \sigma^2(R_t-A) \right). \\
\end{aligned}
\]
Integrate on both sides,
\[
\begin{aligned}
    -b_1(t)\cdot \sinh \left( \sqrt{\frac{\theta\sigma^2}{2\eta}}(T-t) \right) =& \int_t^T \sinh \left( \sqrt{\frac{\theta\sigma^2}{2\eta}}(T-s) \right) \cdot \left( \theta \mu + \theta^2 \sigma^2(R_s-A) \right) \dd s. \\
    \Rightarrow b_1(t) =& -\frac{1}{\sinh \left( \sqrt{\frac{\theta\sigma^2}{2\eta}}(T-t) \right)} \cdot \int_t^T \sinh \left( \sqrt{\frac{\theta\sigma^2}{2\eta}}(T-s) \right) \cdot \left( \theta \mu + \theta^2 \sigma^2(R_s-A) \right) \dd s.
\end{aligned}
\]
Note that now
\[
    v_t = -x_t' = -(x_t-A)' = \frac{b_2}{H}(x_t-A)+\frac{b_1}{2H},
\]
hence,
\[
\begin{aligned}
    \left(\frac{x_t-A}{\sinh \left( \sqrt{\frac{\theta\sigma^2}{2\eta}}(T-t) \right)}\right)' =& \frac{(x_t-A)'\cdot \sinh \left( \sqrt{\frac{\theta\sigma^2}{2\eta}}(T-t) \right)+(x_t-A)\cdot \sqrt{\frac{\theta\sigma^2}{2\eta}}\cdot \cosh \left( \sqrt{\frac{\theta\sigma^2}{2\eta}}(T-t) \right) }{\sinh^2 \left( \sqrt{\frac{\theta\sigma^2}{2\eta}}(T-t) \right)} \\
    =& \frac{\int_t^T \sinh \left( \sqrt{\frac{\theta\sigma^2}{2\eta}}(T-s) \right) \cdot \left( \theta \mu + \theta^2 \sigma^2(R_s-A) \right) \dd s}{2H\sinh^2 \left( \sqrt{\frac{\theta\sigma^2}{2\eta}}(T-t) \right)}.
\end{aligned}
\]
Integrate on both sides then integrate by parts, we have
\[
\begin{aligned}
    &\frac{x_t-A}{\sinh \left( \sqrt{\frac{\theta\sigma^2}{2\eta}}(T-t) \right)} - \frac{x_0-A}{\sinh \left( \sqrt{\frac{\theta\sigma^2}{2\eta}}T \right)} = \int_0^t \int_s^T \frac{\sinh \left( \sqrt{\frac{\theta\sigma^2}{2\eta}}(T-u) \right)}{2H \cdot \sinh^2 \left( \sqrt{\frac{\theta\sigma^2}{2\eta}}(T-s) \right)} \cdot \left( \theta \mu + \theta^2 \sigma^2(R_u-A) \right) \dd u \dd s \\
    =& \frac{1}{\sinh \left( \sqrt{\frac{\theta \sigma^2}{2\eta}}T \right)} \cdot  \int_0^t \frac{1}{\sqrt{2\theta^3\sigma^2\eta}} \sinh \left( \sqrt{\frac{\theta \sigma^2}{2\eta}}s \right) \cdot \left( \theta \mu + \theta^2 \sigma^2(R_s-A) \right) \dd s+\frac{\sinh \left( \sqrt{\frac{\theta \sigma^2}{2\eta}}t \right)}{\sinh \left( \sqrt{\frac{\theta \sigma^2}{2\eta}}T \right)} \cdot \frac{1}{\sinh \left( \sqrt{\frac{\theta \sigma^2}{2\eta}}(T-t) \right)} \\
    & \cdot \int_t^T \frac{1}{\sqrt{2\theta^3\sigma^2\eta}}\cdot \sinh \left( \sqrt{\frac{\theta\sigma^2}{2\eta}}(T-s) \right) \cdot \left( \theta \mu + \theta^2 \sigma^2(R_s-A) \right) \dd s.
\end{aligned}
\]
Denote 
\[
    \frac{\theta^2\sigma^2}{\sqrt{2\theta^3 \sigma^2 \eta}} = \sqrt{\frac{\theta \sigma^2}{2\eta}} =: \kappa,
\]
we can solve that
\[
\begin{aligned}
    x_t =& \frac{\kappa}{\sinh \kappa T} \int_0^T \bigg ( R_s  \sinh (\kappa \min(s,t)) \sinh (\kappa (T-\max(s,t))) \bigg) \ddd s \\
    &+ \left( x_0 -\frac{\mu}{\theta \sigma^2} \right) \cdot \frac{\sinh \kappa(T-t)}{\sinh \kappa T} + A + \left( -A + \frac{\mu}{\theta \sigma^2}\right) \cdot \frac{\sinh \kappa T - \sinh \kappa t}{\sinh \kappa T}.
\end{aligned}
\]
Furthermore, 
\[
\begin{aligned}
    x_t =& \frac{\mu}{\theta \sigma^2} + \left(x_0-\frac{\mu}{\theta \sigma^2}\right)\cdot \frac{\sinh \kappa(T-t)}{\sinh \kappa T} + \left(A-\frac{\mu}{\theta \sigma^2}\right)\cdot \frac{\sinh \kappa t}{\sinh \kappa T} \\
    &+ \frac{\sinh \kappa(T-t)}{\sinh \kappa T} \cdot \int_0^t \kappa \sinh \kappa s R_s \dd s + \frac{\sinh \kappa t}{\sinh \kappa T} \cdot \int_t^T \kappa  \sinh \kappa(T-s) R_s \dd s \\
    =& \frac{\mu}{\theta \sigma^2} + \frac{\sinh \kappa(T-t)}{\sinh \kappa T} \cdot \left( x_0 - \frac{\mu}{\theta \sigma^2} + \int_0^t \kappa \sinh \kappa s R_s \dd s\right) + \frac{\sinh \kappa t}{\sinh \kappa T} \cdot \left( A-\frac{\mu}{\theta \sigma^2} +  \int_t^T \kappa  \sinh \kappa(T-s) R_s \dd s\right)
\end{aligned}
\]

$\hfill \square$

\subsection{Proof to Lemma \ref{lemma:integral}} \label{proof:lemma:integral}

\[
    \begin{aligned}
        & \frac{\kappa}{\sinh \kappa T} \int_0^T \bigg (\sinh (\kappa \min(s,t)) \sinh (\kappa (T-\max(s,t))) \bigg) \dd  s\\
        =&   \frac{\kappa}{\sinh \kappa T} \cdot \int_0^t \sinh \kappa s \dd s \cdot \sinh(\kappa(T-t)) + \frac{\kappa}{\sinh \kappa T} \cdot \int_t^T \sinh \kappa (T-s) \dd s \cdot \sinh\kappa t \\
        =& \frac{\sinh (\kappa(T-t))}{\sinh \kappa T} \cdot (\cosh \kappa t-1) + \frac{\sinh \kappa t}{\sinh \kappa T} \cdot (\cosh \kappa(T-t)-1) \\
        =& \frac{\sinh \kappa T-\sinh \kappa t}{\sinh \kappa T} - \frac{\sinh \kappa (T-t)}{\sinh \kappa T} = \text{TC}_t - \text{IS}_t.
    \end{aligned}
    \]
    $\hfill \square$

\subsection{Proof to Proposition \ref{prop:const_RS}} \label{proof:prop:const_RS}

By Theorem \ref{thm:det_ctrl} and Lemma \ref{lemma:integral},
    \[
    \begin{aligned}
    x_t =& \frac{\kappa}{\sinh \kappa T} \int_0^T \bigg ( R  \sinh (\kappa \min(s,t)) \sinh (\kappa (T-\max(s,t))) \bigg) \ddd s \\
    &+ \left( x_0 -\frac{\mu}{\theta \sigma^2} \right) \cdot \frac{\sinh \kappa(T-t)}{\sinh \kappa T} + A + \left( -A + \frac{\mu}{\theta \sigma^2}\right) \cdot \frac{\sinh \kappa T - \sinh \kappa t}{\sinh \kappa T} \\
    =& \left( x_0-\frac{\mu}{\theta \sigma^2}-R \right)\cdot \text{IS}_t + \left( -A+ \frac{\mu}{\theta \sigma^2}+R \right) \cdot \text{TC}_t + A.
\end{aligned}
\]
$\hfill \square$

\subsection{Proof to Proposition \ref{prop:over_shooting}} \label{proof:prop:over_shooting}
    Note that $R+\frac{\mu}{\theta \sigma^2}$ appears as a whole, so WLOG, we may assume that $\mu \equiv 0$. When the reference strategy is $R$, the optimal strategy is given by
\[
    x_t = A + (R-A) \cdot \text{TC}_t + (x_0-R) \cdot \text{IS}_t,
\]
then
\[
\begin{aligned}
    v_t = -x_t' =& -(R-A) \cdot \text{TC}_t' - (x_0-R) \cdot \text{IS}_t' \\
    =& -(R-A) \cdot \frac{-\kappa \cosh \kappa t}{\sinh \kappa T} - (x_0-R) \cdot \frac{-\kappa \cosh \kappa (T-t)}{\sinh \kappa T},
\end{aligned}
\]
note that $0\le t\le T$, then $\cosh \kappa t \ge 0$ and $\cosh \kappa (T-t) \ge 0$, so when $A\le R\le x_0$, $v_t \ge 0$, which represents that the optimal strategy does not contain overshooting. Note that
\[
    x_t''=-v_t' = (R-A)\cdot \frac{-\kappa^2 \sinh \kappa t}{\sinh \kappa T} + (x_0-R) \cdot \frac{\kappa^2 \sinh \kappa (T-t) }{\sinh \kappa T},
\]
so when $R>x_0>A$, $x_t''=-v_t'\le 0$, which implies that the optimal strategy is a concave function, so
\[
\begin{aligned}
    \min_{0\le t\le T} v_t =& v_0 = -(R-A)\cdot \frac{-\kappa}{\sinh \kappa T} - (x_0-R) \cdot \frac{-\kappa \cosh \kappa T}{\sinh \kappa T}\\
    =& \frac{\kappa}{\sinh \kappa T} \cdot \left[ (1-\cosh \kappa T) R + (x_0\cosh \kappa T-A)\right],
\end{aligned}
\]
so when
\[
    R > \frac{x_0 \cosh \kappa T-A}{\cosh \kappa T -1} = x_0 + \frac{1}{\cosh \kappa T -1} \cdot (x_0-A),
\]
the optimal strategy contains overshooting, and when $x_0<R\le x_0+\frac{1}{\cosh \kappa T -1} \cdot (x_0-A)$, the optimal strategy does not contains overshooting. Similarly, when $R<A<x_0$, $q_t''=-v_t'\ge 0$, which implies that the optimal strategy is a convex function, so
\[
\begin{aligned}
    \min_{0\le t\le T} v_t =& v_T = -(R-A)\cdot \frac{-\kappa \cosh \kappa T }{\sinh \kappa T} - (x_0-R) \cdot \frac{-\kappa}{\sinh \kappa T}\\
    =& \frac{\kappa}{\sinh \kappa T} \cdot \left[ (\cosh \kappa T-1) R - (x_0-A\cosh \kappa T)\right],
\end{aligned}
\]
so when
\[
    R < \frac{A\cosh \kappa T-x_0}{\cosh \kappa T -1} = A - \frac{1}{\cosh \kappa T -1} \cdot (x_0-A),
\]
the optimal strategy contains overshooting.

$\hfill \square$

\subsection{Proof to Proposition \ref{prop:piecewise}} \label{proof:prop:piecewise}

By the results given in Theorem \ref{thm:det_ctrl}, the optimal strategy is a deterministic one. In other words, when $\mu=m_0=0$, 
\[
    \sup_{v\in \mathcal{A}} \mathbb{E} \left[ u\left( \tilde{\Pi} \right) \right] = \sup_{v\in \mathcal{A}_{\text{det}}} \mathbb{E} \left[ u\left( \tilde{\Pi} \right) \right],
\]
where $\mathcal{A}_{\text{det}} = \{ v\in \mathcal{A}: v \text{ is deterministic} \}.$ When $v \in \mathcal{A}_{\text{det}}$, $x_t = x_0 - \int_0^t v_s \dd s$ is also deterministic, hence,
\[
\begin{aligned}
    \mathbb{E} \left[ u\left( \tilde{\Pi} \right) \right] =& 
                 \mathbb{E} \Bigg[ \frac{1}{\theta} \Bigg( 1-\exp\Bigg\{ \theta \beta (x_T-A)^2-\int_0^T \theta \left(-\eta v_t^2-\gamma v_t(x_t-A) \right) \dd t - \int_0^T \theta\sigma (x_t-R_t) \dd W_t \Bigg\} \Bigg)   \Bigg]\\
              =& \frac{1}{\theta} - \frac{1}{\theta} \cdot \exp \Bigg\{\theta \beta (x_T-A)^2-\int_0^T \left[ \theta \left(-\eta v_t^2-\gamma v_t(x_t-A) \right) -\frac{1}{2}\theta^2\sigma^2(x_t-R_t)^2 \right]  \dd t \Bigg\}.
\end{aligned}
\]
Subject to $x_T=A$ (or $\beta \to +\infty)$, $\int_0^T v_t(x_t-A) \dd t = \frac{1}{2}(x_0-A)^2$ is a constant, hence, maximizing the expected utility is equivalent to a typical variational problem
\[
\begin{cases}
\begin{aligned}
    \inf_{v \in \mathcal{A}_{\text{det}}} & \int_0^T \left[\eta v_t^2 +\frac{\theta \sigma^2}{2}(x_t-R_t)^2 \right]  \dd t \\
    \text{s.t. } & x_t = x_0 - \int_0^t v_s \dd s, \  x_T=A,
\end{aligned}
\end{cases}
\]
Note that the objective can be divided into $n$ parts, each with a constant RS:
    \[
        \int_0^T \left( \eta v_t^2 + \frac{\gamma}{2} \sigma^2 (x_t-R_t)^2 \right) \text{d}t = \sum_{k=1}^n \int_{\frac{(k-1)T}{n}}^{\frac{kT}{n}} \left( \eta v_t^2 + \frac{\gamma}{2}\sigma^2 \left(x_t-R^{(k)}\right)^2 \right) \text{d}t.
    \]
    We divide the optimization problem into three steps:
    \begin{itemize}
\item Find the optimal value of $x_{\frac{kT}{n}}^*, 1\le k\le n-1$. We can directly use Theorem \ref{thm:det_ctrl}: suppose the optimal strategy is $x_t^*$, note that $\mu=0$,  
        \[
    \begin{aligned}
    x_t^* =& \frac{\kappa}{\sinh \kappa T} \int_0^T \bigg ( R_s  \sinh (\kappa \min(s,t)) \sinh (\kappa (T-\max(s,t))) \bigg) \ddd s \\
    &+ x_0  \cdot \frac{\sinh \kappa(T-t)}{\sinh \kappa T} + A  -A  \cdot \frac{\sinh \kappa T - \sinh \kappa t}{\sinh \kappa T},
\end{aligned}
\]
where $R$ is the given piece-wise constant RS. For $1\le k\le n-1$,
\[
    \begin{aligned}
    x_{\frac{kT}{n}}^* =& x_0  \cdot \frac{\sinh \kappa(T-\frac{kT}{n})}{\sinh \kappa T} +A  \cdot \frac{\sinh \kappa \frac{kT}{n}}{\sinh \kappa T} +  \frac{\kappa \sinh \kappa \left( T-\frac{kT}{n} \right) }{\sinh \kappa T} \cdot \sum_{i=1}^k R^{(i)}\cdot \int_{\frac{(i-1)T}{n}}^{\frac{iT}{n}}    \sinh (\kappa s)  \dd s \\
    &+ \frac{\kappa \sinh \kappa \frac{kT}{n}}{\sinh \kappa T} \cdot \sum_{i=k+1}^{n} R^{(i)}\cdot  \int_{\frac{(i-1)T}{n}}^{\frac{iT}{n}}   \sinh \left(\kappa \left( T-s \right) \right) \dd s \\
    =& x_0  \cdot \frac{\sinh \kappa(T-\frac{kT}{n})}{\sinh \kappa T} +A  \cdot \frac{\sinh \kappa \frac{kT}{n}}{\sinh \kappa T} \\
    &+  \frac{\sinh \kappa \left( T-\frac{kT}{n} \right) }{\sinh \kappa T} \cdot \sum_{i=1}^k R^{(i)}\cdot \left( \cosh \left( \kappa \frac{iT}{n} \right) - \cosh \left( \kappa \frac{(i-1)T}{n} \right) \right) \\
    &+ \frac{\kappa \sinh \kappa \frac{kT}{n}}{\sinh \kappa T} \cdot \sum_{i=k+1}^{n} R^{(i)}\cdot  \left( \cosh \left( \kappa \frac{(n-i+1)T}{n} \right) - \cosh \left( \kappa \frac{(n-i)T}{n} \right) \right).
\end{aligned}
\]
Denote $R^{(0)}:=x_0$, $R^{(n+1)}:=A$, and
\[
\begin{aligned}
    b_i =\begin{cases}
        1, & i=0,\\
        \cosh \left( \kappa \frac{iT}{n} \right) - \cosh \left( \kappa \frac{(i-1)T}{n} \right), & 1\le i\le n-1,
    \end{cases}
\end{aligned}
\]
we can rewrite
\[
    x_{\frac{kT}{n}}^* = \frac{\sinh \kappa\left(T-\frac{kT}{n}\right)}{\sinh \kappa T} \cdot \sum_{i=0}^k b_i R^{(i)}  + \frac{\sinh \kappa\frac{kT}{n}}{\sinh \kappa T} \cdot \sum_{i=k+1}^{n+1} b_{n-i+1} R^{(i)}, \ 1\le k\le n-1.
\]
Furthermore, we denote the optimal values $a_k :=  x_{\frac{kT}{n}}^*, 1\le k\le n-1$. Note that the sum of coefficients in $a_k$ is
\[
\begin{aligned}
    & \frac{\sinh \kappa\left(T-\frac{kT}{n}\right)}{\sinh \kappa T} \cdot \sum_{i=0}^k b_i   + \frac{\sinh \kappa\frac{kT}{n}}{\sinh \kappa T} \cdot \sum_{i=k+1}^{n+1} b_{n-i+1} \\
    =& \frac{\sinh \kappa\left(T-\frac{kT}{n}\right)}{\sinh \kappa T} \cdot \cosh \kappa\frac{kT}{n}   + \frac{\sinh \kappa\frac{kT}{n}}{\sinh \kappa T} \cdot \cosh \kappa \frac{(n-k)T}{n} = 1,
\end{aligned}
\]
so $a_k$ is a weighted average of $R^{(i)}$'s.
\item Note that we already know the value of optimal strategy $x_{\frac{kT}{n}}$, so the optimization problem is equivalent to the following optimization problem with extra constraints:
\[
\begin{cases}
\begin{aligned}
    \inf_{v \in \mathcal{A}_{\text{det}}} & \int_0^T \left[\eta v_t^2 +\frac{\theta \sigma^2}{2}(x_t-R_t)^2 \right]  \dd t = \inf_{v \in \mathcal{A}_{\text{det}}} \sum_{k=1}^n \int_{\frac{(k-1)T}{n}}^{\frac{kT}{n}} \left( \eta v_t^2 + \frac{\gamma}{2}\sigma^2 \left(x_t-R^{(k)}\right)^2 \right) \\
    \text{s.t. } & x_t = x_0 - \int_0^t v_s \dd s, \ x_T=A, \\
    & x_{\frac{kT}{n}} = a_k, \ 1\le k\le n-1.
\end{aligned}
\end{cases}
\]
Denote $a_0=x_0$ and $a_n=A$, inheriting from the spirit of dynamic programming, we can further divide the optimization problem into $n$ parts: for $1\le k \le n$, where the $k$-th sub-problem is given by
\[
\begin{cases}
\begin{aligned}
    \inf_{v \in \mathcal{A}_{\text{det}}} & \int_{\frac{(k-1)T}{n}}^{\frac{kT}{n}}\left( \eta v_t^2 + \frac{\theta \sigma^2}{2} \left(x_t-R^{(k)}\right)^2 \right) \text{d}t \\
    \text{s.t. } & x_t = a_{k-1} - \int_\frac{(k-1)T}{n}^t v_s \dd s, \ \frac{(k-1)T}{n}\le t\le \frac{kT}{n}, \  x_\frac{kT}{n}=a_k,
\end{aligned}
\end{cases}
\]
which is a constant RS problem with time period $\left[ \frac{(k-1)T}{n}, \frac{kT}{n} \right]$, of which the optimal strategy is given by
\[
        x_t^{(k)} = a_k + \left( R^{(k)} -a_k \right) \, \text{TC}_{t-\frac{(k-1)T}{n}} + \left( a_{k-1}-R^{(k)}\right) \, \text{IS}_{t-\frac{(k-1)T}{n}}, \frac{(k-1)T}{n} \le t\le \frac{kT}{n},
    \]
    where $\kappa:=\sqrt{\frac{\theta \sigma^2}{2\eta}}$ and
    \[
    {\rm IS}_t := \frac{\sinh \kappa \left(\frac{T}{n}-t\right)}{\sinh \kappa \frac{T}{n}}, \ \ {\rm TC}_t := \frac{\sinh \kappa \frac{T}{n}-\sinh \kappa t}{\sinh \kappa \frac{T}{n}}.
    \]
\item Connect $n$ optimal strategies to obtain the optimal strategy of the original optimization problem. Denote
\[
    x_t^{**} = \sum_{k=1}^n x_t^{(k)} \, \mathbb{1}_{\{ \frac{(k-1)T}{n}\le t<\frac{kT}{n} \}} + A \, \mathbb{1}_{\{t=T\}},
\]
then by the results of constant reference strategies, for any strategy $x_t$,
\[
\begin{aligned}
    \int_0^T \left[\eta v_t^2 +\frac{\theta \sigma^2}{2}(x_t-R_t)^2 \right]  \dd t =& \sum_{k=1}^n \int_{\frac{(k-1)T}{n}}^{\frac{kT}{n}} \left( \eta v_t^2 + \frac{\gamma}{2}\sigma^2 \left(x_t-R^{(k)}\right)^2 \right) \\
    \ge & \sum_{k=1}^n \int_{\frac{(k-1)T}{n}}^{\frac{kT}{n}} \left( \eta \left(-\dot{x}_t^{(k)}\right)^2 + \frac{\gamma}{2}\sigma^2 \left(x_t^{(k)}-R^{(k)}\right)^2 \right) \\
    =& \int_0^T \left[\eta \left(-\dot{x}_t^{**}\right)^2 +\frac{\theta \sigma^2}{2}(x_t^{**}-R_t)^2 \right]  \dd t,
\end{aligned}
\]
where implies that the optimal value is obtained when $x_t=x_t^{**}$, in other words, $x_t^{**}$ is the optimal strategy.
    \end{itemize}
$\hfill \square$

\subsection{Proof to Theorem \ref{thm:approx}} \label{proof:thm:approx}

According to Theorem \ref{thm:det_ctrl}, the optimal strategies $x_t$ and $\widetilde{x}_t$ can be given by
    \[
    \begin{aligned}
    x_t =& \frac{\kappa}{\sinh \kappa T} \int_0^T \bigg ( R_s \sinh (\kappa \min(s,t)) \sinh (\kappa (T-\max(s,t))) \bigg) \ddd s \\
    &+ \left(x_0-\frac{\mu}{\theta \sigma^2}\right) \cdot \frac{\sinh \kappa(T-t)}{\sinh \kappa T} + A + \left(-A+\frac{\mu}{\theta \sigma^2}\right) \cdot \frac{\sinh \kappa T - \sinh \kappa t}{\sinh \kappa T},
\end{aligned}
\]
and
 \[
    \begin{aligned}
    \widetilde{x}_t =& \frac{\kappa}{\sinh \kappa T} \int_0^T \bigg ( \widetilde{R}_t \sinh (\kappa \min(s,t)) \sinh (\kappa (T-\max(s,t))) \bigg) \ddd s \\
    &+ \left(x_0-\frac{\mu}{\theta \sigma^2}\right) \cdot \frac{\sinh \kappa(T-t)}{\sinh \kappa T} + A + \left(-A+\frac{\mu}{\theta \sigma^2}\right) \cdot \frac{\sinh \kappa T - \sinh \kappa t}{\sinh \kappa T},
\end{aligned}
\]
where $x_t, \widetilde{x}_t$ are the optimal strategy with respect to reference strategies $R_t$ and $\widetilde{R}_t$ respectively. Hence, by Jensen's inequality,
\[
\begin{aligned}
    \left(x_t-\widetilde{x}_t\right)^2 =& \frac{\kappa^2}{\sinh^2 \kappa T} \left(\int_0^T \bigg ( \left(R_s^{(1)}-R_s^{(2)}\right) \sinh (\kappa \min(s,t)) \sinh (\kappa (T-\max(s,t))) \bigg) \ddd s\right)^2 \\
    \le & \frac{\kappa^2 }{\sinh^2 \kappa T} \int_0^T \left(R_s^{(1)}-R_s^{(2)}\right)^2 \ddd s \cdot \int_0^T \sinh^2(\kappa \min(s,t))\sinh^2(\kappa (T-\max(s,t)))\ddd s \\
    <& \frac{\kappa^2}{\sinh^2 \kappa T} \cdot \varepsilon \cdot \frac{1}{4\kappa}\left( \sinh^2(\kappa t)\left( \sinh(2\kappa(T-t))-2\kappa(T-t) \right) + \sinh^2(\kappa (T-t))\left( \sinh(2\kappa t)-2\kappa t \right) \right) \\
    \le & \frac{\kappa^2}{\sinh^2 \kappa T} \cdot \varepsilon \cdot \frac{1}{2\kappa} \sinh^2 \left( \kappa \frac{T}{2} \right) \left( \sinh \kappa T -\kappa T \right) = \frac{\kappa \left( \sinh \kappa T -\kappa T \right)}{4 \left(\cosh\kappa T + 1 \right)} \cdot \varepsilon < \frac{\kappa \varepsilon}{4}\\
\end{aligned}
\]
which implies the fact that when $\varepsilon \to 0$, $\widetilde{x}_t\to x_t$. So we may use the optimal strategy by a piece-wise constant RS to approximate the general RS, with small error
\[
    \left|x_t-\widetilde{x}_t\right| < \frac{1}{2} \sqrt{\kappa \varepsilon}.
\]

$\hfill \square$

\newpage
\bibliographystyle{IEEEtran}
\bibliography{ref}

\begin{thebibliography}{10}
\providecommand{\url}[1]{#1}
\csname url@samestyle\endcsname
\providecommand{\newblock}{\relax}
\providecommand{\bibinfo}[2]{#2}
\providecommand{\BIBentrySTDinterwordspacing}{\spaceskip=0pt\relax}
\providecommand{\BIBentryALTinterwordstretchfactor}{4}
\providecommand{\BIBentryALTinterwordspacing}{\spaceskip=\fontdimen2\font plus
\BIBentryALTinterwordstretchfactor\fontdimen3\font minus \fontdimen4\font\relax}
\providecommand{\BIBforeignlanguage}[2]{{%
\expandafter\ifx\csname l@#1\endcsname\relax
\typeout{** WARNING: IEEEtran.bst: No hyphenation pattern has been}%
\typeout{** loaded for the language `#1'. Using the pattern for}%
\typeout{** the default language instead.}%
\else
\language=\csname l@#1\endcsname
\fi
#2}}
\providecommand{\BIBdecl}{\relax}
\BIBdecl

\bibitem{almgren2001optimal}
R.~Almgren and N.~Chriss, ``Optimal execution of portfolio transactions,'' \emph{Journal of Risk}, vol.~3, pp. 5--40, 2001.

\bibitem{cfa2017cfa}
C.~Institute, \emph{CFA Program Curriculum 2018 Level II}.\hskip 1em plus 0.5em minus 0.4em\relax John Wiley \& Sons, 2017.

\bibitem{gueanttarget}
\BIBentryALTinterwordspacing
O.~Gu{\'e}ant, ``Target close execution strategies.'' [Online]. Available: \url{https://www.oliviergueant.com/uploads/4/3/0/9/4309511/targetclose.pdf}
\BIBentrySTDinterwordspacing

\bibitem{almgren1999value}
R.~Almgren and N.~Chriss, ``Value under liquidation,'' \emph{Risk}, vol.~12, no.~12, pp. 61--63, 1999.

\bibitem{doi:10.1080/135048602100056}
R.~F. Almgren, ``Optimal execution with nonlinear impact functions and trading-enhanced risk,'' \emph{Applied Mathematical Finance}, vol.~10, no.~1, pp. 1--18, 2003.

\bibitem{bertsimas1998optimal}
D.~Bertsimas and A.~W. Lo, ``Optimal control of execution costs,'' \emph{Journal of Financial Markets}, vol.~1, no.~1, pp. 1--50, 1998.

\bibitem{gatheral2010no}
J.~Gatheral, ``No-dynamic-arbitrage and market impact,'' \emph{Quantitative finance}, vol.~10, no.~7, pp. 749--759, 2010.

\bibitem{gatheral2012transient}
J.~Gatheral, A.~Schied, and A.~Slynko, ``Transient linear price impact and fredholm integral equations,'' \emph{Mathematical Finance: An International Journal of Mathematics, Statistics and Financial Economics}, vol.~22, no.~3, pp. 445--474, 2012.

\bibitem{tse2013comparison}
S.~Tse, P.~Forsyth, J.~Kennedy, and H.~Windcliff, ``Comparison between the mean-variance optimal and the mean-quadratic-variation optimal trading strategies,'' \emph{Applied Mathematical Finance}, vol.~20, no.~5, pp. 415--449, 2013.

\bibitem{cheng2019optimal}
X.~Cheng, M.~Di~Giacinto, and T.-H. Wang, ``Optimal execution with dynamic risk adjustment,'' \emph{Journal of the Operational Research Society}, vol.~70, no.~10, pp. 1662--1677, 2019.

\bibitem{frei2015optimal}
C.~Frei and N.~Westray, ``Optimal execution of a vwap order: a stochastic control approach,'' \emph{Mathematical Finance}, vol.~25, no.~3, pp. 612--639, 2015.

\bibitem{cartea2016closed}
{\'A}.~Cartea and S.~Jaimungal, ``A closed-form execution strategy to target volume weighted average price,'' \emph{SIAM Journal on Financial Mathematics}, vol.~7, no.~1, pp. 760--785, 2016.

\bibitem{gueant2014vwap}
O.~Gu{\'e}ant and G.~Royer, ``Vwap execution and guaranteed vwap,'' \emph{SIAM Journal on Financial Mathematics}, vol.~5, no.~1, pp. 445--471, 2014.

\bibitem{almgren2007adaptive}
R.~Almgren and J.~Lorenz, ``Adaptive arrival price,'' \emph{Trading}, vol. 2007, no.~1, pp. 59--66, 2007.

\bibitem{frei2018optimal}
C.~Frei and N.~Westray, ``Optimal execution in hong kong given a market-on-close benchmark,'' \emph{Quantitative Finance}, vol.~18, no.~4, pp. 655--671, 2018.

\bibitem{gueant2015accelerated}
O.~Gu{\'e}ant, J.~Pu, and G.~Royer, ``Accelerated share repurchase: pricing and execution strategy,'' \emph{International Journal of Theoretical and Applied Finance}, vol.~18, no.~03, p. 1550019, 2015.

\bibitem{bargeron2011accelerated}
L.~Bargeron, M.~Kulchania, and S.~Thomas, ``Accelerated share repurchases,'' \emph{Journal of Financial Economics}, vol. 101, no.~1, pp. 69--89, 2011.

\bibitem{cheng2017optimal}
X.~Cheng, M.~Di~Giacinto, and T.-H. Wang, ``Optimal execution with uncertain order fills in almgren--chriss framework,'' \emph{Quantitative Finance}, vol.~17, no.~1, pp. 55--69, 2017.

\bibitem{carmona2023optimal}
R.~Carmona and L.~Leal, ``Optimal execution with quadratic variation inventories,'' \emph{SIAM Journal on Financial Mathematics}, vol.~14, no.~3, pp. 751--776, 2023.

\bibitem{nutz2023unwinding}
M.~Nutz, K.~Webster, and L.~Zhao, ``Unwinding stochastic order flow: When to warehouse trades,'' \emph{arXiv preprint arXiv:2310.14144}, 2023.

\bibitem{cartea2022brokers}
{\'A}.~Cartea and L.~S{\'a}nchez-Betancourt, ``Brokers and informed traders: dealing with toxic flow and extracting trading signals,'' \emph{Available at SSRN 4265814}, 2022.

\bibitem{yong2012stochastic}
J.~Yong and X.~Y. Zhou, \emph{Stochastic controls: Hamiltonian systems and HJB equations}.\hskip 1em plus 0.5em minus 0.4em\relax Springer Science \& Business Media, 2012, vol.~43.

\bibitem{duncan2013linear}
T.~E. Duncan, ``Linear-exponential-quadratic gaussian control,'' \emph{IEEE Transactions on Automatic Control}, vol.~58, no.~11, pp. 2910--2911, 2013.

\bibitem{lim2005new}
A.~E. Lim and X.~Y. Zhou, ``A new risk-sensitive maximum principle,'' \emph{IEEE transactions on automatic control}, vol.~50, no.~7, pp. 958--966, 2005.

\bibitem{carmona2013self}
R.~Carmona and K.~Webster, ``\BIBforeignlanguage{English (US)}{The self-financing equation in limit order book markets},'' \emph{\BIBforeignlanguage{English (US)}{Finance and Stochastics}}, vol.~23, no.~3, pp. 729--759, Jul. 2019.

\bibitem{mazzolo2017constraint}
A.~Mazzolo, ``{Constraint Ornstein-Uhlenbeck bridges},'' \emph{Journal of Mathematical Physics}, vol.~58, no.~9, p. 093302, 09 2017.

\bibitem{gueant2016financial}
O.~Gu{\'e}ant, \emph{The Financial Mathematics of Market Liquidity: From optimal execution to market making}.\hskip 1em plus 0.5em minus 0.4em\relax CRC Press, 2016, vol.~33.

\bibitem{merton1975optimum}
R.~C. Merton, ``Optimum consumption and portfolio rules in a continuous-time model,'' in \emph{Stochastic optimization models in finance}.\hskip 1em plus 0.5em minus 0.4em\relax Elsevier, 1975, pp. 621--661.

\end{thebibliography}
\addcontentsline{toc}{section}{Reference} 

\end{document}